\documentclass[12pt,preprint]{aastex}

 \input{psfig}
 \input{epsf}

\def\lya{\ifmmode {\rm Ly}\alpha~ \else Ly$\alpha$~\fi}
\def\lyb{\ifmmode {\rm Ly}\beta~ \else Ly$\beta$~\fi}
\def\lyg{\ifmmode {\rm Ly}\gamma~ \else Ly$\gamma$~\fi}
\def\civ{\ifmmode {\rm C}\,{\sc iv}~ \else C\,{\sc iv}~\fi}
\def\civn{\ifmmode {\rm C}\,{\sc iv}~ \else C\,{\sc iv}\fi}
\def\cvn{\ifmmode {\rm C}\,{\sc v}~ \else C\,{\sc v}\fi}
\def\cvin{\ifmmode {\rm C}\,{\sc vi}~ \else C\,{\sc vi}\fi}
\def\nvn{\ifmmode {\rm N}\,{\sc v}~ \else N\,{\sc v}\fi}
\def\nvin{\ifmmode {\rm N}\,{\sc vi}~ \else N\,{\sc vi}\fi}
\def\nviin{\ifmmode {\rm N}\,{\sc vii}~ \else N\,{\sc vii}\fi}

\def\oi{{{\rm O}\,\hbox{{\sc i}}~}}
\def\oin{{{\rm O}\,\hbox{{\sc i}}}}

\def\ovi{{{\rm O}\,\hbox{{\sc vi}}~}}
\def\ovii{{{\rm O}\,\hbox{{\sc vii}}~}}
\def\oviii{{{\rm O}\,\hbox{{\sc viii}}~}}
\def\ovn{{{\rm O}\,\hbox{{\sc v}}}}
\def\ovin{{{\rm O}\,\hbox{{\sc vi}}}}
\def\oviin{{{\rm O}\,\hbox{{\sc vii}}}}
\def\oviiin{{{\rm O}\,\hbox{{\sc viii}}}}
\def\neix{\ifmmode {\rm Ne}\,{\sc ix}~ \else Ne\,{\sc ix}~\fi}
\def\nex{\ifmmode {\rm Ne}\,{\sc x}~ \else Ne\,{\sc x}~\fi}
\def\hi{\ifmmode {\rm H}\,{\sc i}~ \else H\,{\sc i}~\fi}
\def\kms{\rm\,km\,s^{-1}}
\def\hubunits{\rm\,km\,s^{-1}\,Mpc^{-1}}
\def\K{\,{\rm K}}
\def\cm{{\rm cm}}
\def\chandra {{\it Chandra}~}

\def\etal   {{\it et~al.}~}

\slugcomment{Submitted to ApJ}

\shorttitle{X-ray Forest}
\shortauthors{Mathur \etal}

\begin{document}


\title{Tracing the Warm-Hot Intergalactic Medium at Low Redshift: \\
X-ray Forest Observations Towards H1821+643}


\author{Smita Mathur\altaffilmark{1}, 
David H. Weinberg\altaffilmark{1} \&  
Xuelei Chen\altaffilmark{2,3} 
}

\altaffiltext{1}{Astronomy Department, The Ohio State University, 140
West 18th Avenue, Columbus, OH 43210; smita,
dhw@astronomy.ohio-state.edu}
\altaffiltext{2}{Physics Department, The Ohio State University, 174
West 18th Avenue, Columbus, OH 43210}
\altaffiltext{3}{Institute for Theoretical Physics, U.C. Santa
Barbara, Santa Barbara, CA 93106; xuelei@itp.ucsb.edu}


\begin{abstract}
We present a high resolution ($\lambda/\Delta\lambda\approx 500$) X-ray
spectrum of the bright quasar H1821+643 ($z=0.297$), obtained in a 470
ksec observation with the \chandra X-ray Observatory.  
We search for X-ray absorption by highly ionized metal species, \ovii and
\oviii in particular, at the redshifts of the six intervening \ovi 
absorption systems known from UV studies.  We detect
features with $\ga 2\sigma$ significance at the predicted
\ovii and \oviii wavelengths of one \ovi system, at the \ovii
wavelength of a second, and at the \neix wavelength of a third. We
find two additional features of comparable strength (one \ovii and one
\oviiin) within $1000\;\kms$ of \ovi redshifts.  The $1\sigma$ constraints
on the relative abundances of different species imply significant
variations from system to system in $f(\ovin)$, the fraction
of oxygen in the \ovi state. The constraints 
in the two detected \ovi systems imply gas overdensities lower than
the values $\delta \ga 100$ expected in virialized systems, suggesting
that the absorption arises in lower density, filamentary structures.
At the $2\sigma$ level, however, the
physical constraints are weak, though all of the systems must have
temperature $T < 10^6\K$ to be consistent with upper limits on \oviin.
If we treat our $2\sigma$ detections of known \ovi systems 
as real, but assume minimal \ovii
and \oviii in the other systems, we estimate
$[f(\ovin)+f(\oviin)+f(\oviiin)]/f(\ovin)=32 \pm 9$ for the average
ratio of all highly ionized oxygen species to \ovin.  Combined with
estimates of the total column density of \ovi absorption per unit
redshift, this ratio implies that the total baryon fraction associated
with detected \ovi absorbers is $\Omega_b(\ovin)\sim 0.03
h_{70}^{-1}$, a substantial fraction of the baryon density predicted
by big bang nucleosynthesis, and larger than that associated with stars
or with gas detected in 21cm or X-ray emission. Because of the limited S/N of 
the detections, these results must be treated with caution.  Nonetheless,
the combination of the \ovi data with these X-ray forest
measurements provides the most direct evidence to date for the
pervasive, moderate density, shock-heated intergalactic medium
predicted by leading cosmological scenarios.  The high inferred incidence
of relatively strong \ovii and \oviii absorption implies that some
regions of this medium are enriched to a level substantially
above $[{\rm O/H}]=-1$.
\end{abstract}




\section{Introduction}
\label{sec:intro}

Much of our knowledge of the intergalactic medium (IGM) comes from the
rest-frame UV line absorption that it imprints
on the spectra of background quasars: the \lya forest of neutral
hydrogen and associated metal lines such as \civ and \ovi.
The \lya forest is produced mainly by diffuse, photoionized 
gas at temperatures $T\sim 10^4\K$, and it appears to trace the main
reservoir of cosmic baryons at high redshift
\citep{rauch95,hernquist96,rauch97,weinberg97}.
The \lya forest thins out at low redshift, 
and cosmological simulations predict that 
the continuing process of structure formation heats
a substantial fraction of intergalactic gas to temperatures where it 
produces little hydrogen \lya absorption, and where the dominant
ionization stages of heavier elements have absorption transitions
at X-ray wavelengths rather than UV.  Hot, dense gas in the central
regions of galaxy clusters and groups can be detected by its
X-ray emission, but most of the shock-heated gas resides in the 
outskirts of virialized halos and in lower density, filamentary
structures, making its continuum emission weak.
One of the few prospects for detecting this low density,
shock-heated gas is via the ``X-ray forest'' of absorption
lines it should produce in quasar spectra
\citep{hellsten98,perna98,fang00,chen02,fang02a}.
\cite{nicastro02} find strong evidence for absorption by
highly ionized oxygen and neon at $z\approx 0$, but there
have been no clear detections to date of intergalactic X-ray
absorption originating beyond the Local Group.
%{\bf (Any other searches we should reference?)}
This paper describes a search for X-ray forest absorption
towards H1821+643 ($z=0.297$) using a high resolution
($\lambda/\Delta\lambda \approx 500$) spectrum obtained in
a 470 ksec observation with the \chandra X-ray Observatory.

The baryon density implied by big bang nucleosynthesis 
and the estimated primordial deuterium abundance,
$\Omega_{\rm BBN} \approx 0.04 h_{70}^{-2}$ 
(\citealt{burles97,burles98}; here
$h_{70} \equiv H_0/70\;\hubunits$),
exceeds the density of baryons in known stars and X-ray emitting
gas by roughly an order of magnitude \citep{fukugita98}.
The lower density regime of the ``warm-hot intergalactic medium''
(WHIM, a term coined by \citeauthor{cen99} [\citeyear{cen99}] to refer to
gas in the temperature range $10^5-10^7\K$) could constitute a
major fraction of the ``missing'' low redshift baryons.  Hydrodynamic
cosmological simulations predict that 30-50\% of the baryons reside
in this phase at $z=0$ (\citealt{cen99}; \citealt{dave99}, \citeyear{dave01}).
HST and FUSE detections of \ovi ($\lambda\lambda 1032,
1038$\AA) absorption lines towards H1821+643 and PG0953+415 
(\citealt{tripp00}, hereafter TSJ; \citealt{ts00,oegerle00,savage02})
offer a tantalizing hint of this baryon reservoir.
Adopting conservative assumptions of [O/H]$=-1$ and an \ovi
ionization fraction $f$(\ovin)=0.2 (which is close to the maximum
in photo- or collisional ionization), TSJ conclude that the gas
associated with these weak \ovi absorbers accounts for $\Omega_b
\approx 0.004 h_{70}^{-1}$ of the cosmic baryon budget, comparable in
total mass to all other known low redshift components combined.
They remark, moreover, that $f(\ovin)$ could plausibly be much lower
than 0.2, implying a substantially higher baryon fraction.
For most reasonable assumptions
about the physical conditions of this absorbing gas, the dominant
ionization state should be \ovii or \oviiin.
Gas hotter than $T\sim 5\times 10^5\K$ would have
very low $f(\ovin)$, so it might produce \ovii or \oviii absorption
with no detectable \ovin.

We chose H1821+643 for our X-ray forest search in part because it is
one of the brightest X-ray quasar at moderate redshift, and in part because 
it has been studied carefully for intervening \ovi absorption
(TSJ; \citealt{oegerle00}).  Theoretical expectations for
X-ray forest absorption 
depend strongly on the assumed metallicity distribution of the
moderate density, shock-heated IGM, a subject on which there is
little empirical guidance.  However, for metallicities 
$Z \sim 0.1-0.3 Z_\odot$, simulations and analytic calculations
predict few \ovii or \oviii lines above column density 
$N\sim 2\times 10^{15}\cm^{-2}$
\citep{hellsten98,perna98,fang00,chen02,fang02a}, and weaker
absorption by other elements, so even long exposures with 
\chandra or {\it XMM-Newton} are unlikely to detect absorption
at high signal-to-noise ratio.  Searching at redshifts with known
\ovi absorption allows one to adopt a lower effective threshold
for significant detection, greatly increasing the odds of success.
Furthermore, detections of or upper limits on X-ray absorption
provide new constraints on the physical conditions of the \ovi absorbers.

We chose the duration of our spectroscopic observation so that if
$[f(\ovin)+f(\oviin)+f(\oviiin)]/f(\ovin) \sim 30$, a physically
plausible ratio that would imply that \ovi absorbers contain a large
fraction of the ``missing'' low redshift baryons, then we would
clearly detect the strongest of the TSJ systems, and we would detect
the next two strongest systems at the $2-3\sigma$ level.  As we will
show below, we find only a $\sim 1\sigma$ signal from the strongest
TSJ system, but we obtain $\approx 2\sigma$ detections of \ovii from
the next strongest system and of \ovii and \oviii from the \ovi system
discovered by \cite{oegerle00}.  We also find features that could
correspond to oxygen absorption at other redshifts close to the
detected systems (within 1000$\kms$), and a possible signature of
\neix absorption from one system.  
The best estimate of the mean $[f(\ovin)+f(\oviin)+f(\oviiin)]/f(\ovin)$
from our data is $\sim 30$, but the moderate statistical
significance of our detections and the implied variation in $f(\ovin)$
from system to system leave the interpretation of our results more
ambiguous than we would like.
Nonetheless, this observation provides the strongest evidence to date
for the moderate density, shock-heated IGM predicted by theoretical
models, and the detections and upper limits have important
implications for the physical state of the \ovi absorption systems and
for their connection to the missing baryons of the low redshift
universe.  We proceed to a description of the observations, data
reduction, and absorption line analysis, then to a discussion of these
implications.

\section{Observations and Data Reduction}
\label{sec:observations}

We observed H1821+643 with \chandra between 2001 January 17 and
January 24, for a total exposure time of 470 ksec. The observation was
divided into four different parts: ObsIDs 2186, 2418, 2310 and
2311. We used the low energy transmission grating (LETG,
\citealt{brinkman97}) with the advanced CCD imaging camera for
spectroscopy (ACIS-S, C. R. Canizares \etal in preparation).  Our
primary interest is in \ovii and \oviii lines with rest wavelengths of
21.602\AA\ and 18.969\AA, respectively, and thus in the observed
wavelength range 18.9--28\AA\ that runs from \oviii at $z=0$
to \ovii at the quasar redshift $z=0.297$.
Since the lines are expected to be low S/N, it is
important to optimize the instrument response in this interval.  To
avoid the S3/S4 chip gap between 21.2 and 21.8 \AA, the aim point was
offset by 2$^{\prime}$ in +Y direction. This offset moves the chip gap
to 27.9--28.5\AA, so the entire first order spectrum in the
wavelengths of interest is now on S3.  Since the telescope focus is on
S3, this choice enhances the effective area around 25 \AA.  To
ameliorate the effects of CTI for spectral order sorting, a SIM-Z
offset of -8mm was chosen. Accordingly, an ACIS-S subarray with rows
49--256 was used.

The shift of aimpoint, essential for optimizing wavelength coverage
and effective area in the most interesting part of the spectrum,
had the additional effect of placing the zeroth-order grating image
close to the S2/S3 chip gap. This is not a concern from the
scientific point of view, since the X-ray forest search relies on the high
resolution, dispersed spectrum. However, the standard \chandra
processing software (\chandra Interactive Analysis of Observations,
CIAO, Elvis et al.\ in preparation) failed to process the
data properly as a result of this non-standard aim-point. In
particular, the CIAO tool {\bf tg\_resolve\_events}, which assigns a
wavelength and an order to each observed photon, failed. This problem
was fixed in CIAO version 2.2, and the data were reprocessed to
recover all of the observed photons.  We used CIAO version 2.2 in all the
data reduction and analysis.

We extracted spectra from each of the four observations
separately,\footnote{See http://cxc.harvard.edu/ciao/documents\_
threads\_ gspec.html} and $+1$ and $-1$ orders were co-added. The
four resulting spectra were then co-added to construct the full first-order
spectrum. Errors in observed counts were appropriately propagated. The
net observed count rate is 0.428 counts per second for the combined
first-order spectrum, with $2\times 10^5$ total net counts.  The
unbinned spectrum has 8192 wavelength channels with $\Delta\lambda=0.0125$ \AA.
The full width at half maximum of the line response function is about
0.05\AA, increasing to larger values longward of about 50\AA. 
The spectral resolution is therefore
$\lambda/\Delta \lambda \approx 500$ at 25\AA.  The
data span the wavelength range from 1 to 60 \AA, but there is
essentially no detectable flux longward of $\sim 40$\AA.
The understanding of the absolute wavelength calibration of \chandra
spectra is continually evolving,\footnote{See
http://cxc.harvard.edu/cal/calreview/}
but the best current evidence is that the measured wavelengths
near 25\AA\ are offset with respect to true wavelengths by an
amount $0 \leq \Delta\lambda \leq 0.025$\AA. 
While we could add a 0.0125\AA\ offset to all measured
wavelengths and thereby make the calibration uncertainty symmetric
about zero, we have chosen instead to keep the standard wavelength
scale and to consider an observed feature ``matched'' to a predicted
feature at wavelength $\lambda_p$ if it lies between $\lambda_p$
and $\lambda_p+0.025$\AA.

\section{Analysis}
\label{sec:analysis}

In order to detect and measure absorption lines, we first need an
accurate model of the unabsorbed continuum.  The continuum properties of 
H1821+643 have been discussed by \cite{fang02b}, so we will not 
examine them in detail here.  We obtain a continuum for purposes
of line detection by fitting a smooth model to the coarsely binned
first-order spectrum.  The model that we fit is
\begin{equation}
f(E) = K \left(\frac{E}{1\;{\rm keV}}\right)^{-\Gamma} 
         \exp\left[-N_H \sigma(E)\right],
\label{eqn:cont}
\end{equation}
representing power-law emission absorbed by Galactic gas with hydrogen column
density $N_H$ and bound-free absorption cross-section $\sigma(E)$.
However, we found that a single power law did not provide an adequate
fit over the entire range of the data, and we therefore carried out
two independent fits in the range 0.3--0.7 keV (17.7--41.3\AA) and
0.6--1 keV (12.4--20.7\AA).  We use the first fit to search for
lines in the \ovii and \oviii region, and the second to search for
lines in the \neix and \nex region.  Obtaining an adequate representation
of the data further requires that we treat the column density $N_H$
as a free parameter rather than adopting the standard value of
$3.9\times 10^{20}\cm^{-2}$ \citep{lockman95}; in effect, we
use $N_H$ as a parameter to describe departures from a power law, whether 
intrinsic to the quasar or caused by the Galactic ISM.
The fit parameters are: 
in the 0.3--0.7 keV region, 
$K= 0.0030$ photons keV$^{-1}$ cm$^{-2}$ s$^{-1}$, $\Gamma=2.28$, 
$N_H=5.67\times 10^{20}$ cm$^{-2}$;
in the 0.6--1 keV region,
$K= 0.0035$ photons keV$^{-1}$ cm$^{-2}$ s$^{-1}$, $\Gamma=2.08$, 
$N_H=3.69\times 10^{20}$ cm$^{-2}$.
Both of these fits have $\chi^2/\hbox{d.o.f.} \sim 1$, but visual inspection
reveals some excess flux between 24 and 25\AA\ (see Figure~2, top),
which we have chosen to model as two Gaussian ``emission lines'' 
centered at 24.1516\AA\ and 24.5629\AA, with FWHM of 0.3035\AA\ and 
0.120844\AA, respectively.  The amplitudes of these features
(i.e., the integrals under the Gaussians) are $1.795\times 10^{-5}$
and $9.601\times 10^{-6}$ photons cm$^{-2}$ s$^{-1}$, respectively.
The nature of these ``emission lines'' is unclear.
We analyzed several data sets from the \chandra archive that had
similar LETG/ACIS-S instrument set-ups to look for possible systematic
calibration problems, but these do not show signs of excess flux
at these wavelengths.  We expect, therefore, that this flux
originates either from the quasar itself or from the surrounding cluster.
A physical characterization is not essential to our purposes here,
since we only need to define a smooth continuum that fits our data.

Our final continuum model, therefore, consists of one of the two
absorbed power laws described above (depending on the wavelength region),
with the additional Gaussian emission lines in the 24--25\AA\ region,
all modulated by the instrument response, which we fold in by
creating an ``arf'' file generated for our observational set-up.
We began our search for potential X-ray forest systems by examining
the residuals between the data and this smooth continuum model, 
looking for correlations between minima of the residual spectrum
and the absorption wavelengths expected on the basis of the known
\ovi systems.  Figure~1 shows the residual spectrum in the 21--28\AA\
region, where the associated \ovii and \oviii lines fall.
The spectrum in the upper panel has been convolved with a beta
function of FWHM=0.05\AA, which represents the line response function
(LRF) of the instrument at these wavelengths.  This smoothing
suppresses uncorrelated noise fluctuations relative to features that
have the expected shape of true, unresolved absorption (or emission)
lines, and it should therefore enhance the contrast of physical absorption
features relative to photon noise.  However, LRF convolution systematically
broadens features and reduces their central depth (while preserving their
equivalent width), so in the lower panel we show the residual spectrum smoothed
instead with a Savitsky-Golay (S-G) filter 
(see \citealt{press92}, section 14.8).  This type of filter approximately
preserves the sharpness and depth of features, but it is less effective at 
suppressing noise.  The two approaches have complementary advantages,
and in practice the two residual spectra show significant features
at nearly the same locations.

The strong line at $\lambda \approx 23.5$\AA\ is Galactic \oin,
and at somewhat shorter wavelengths, the Galactic O-K edge
and features of the instrument response cause rapid variation of
the model continuum (see Figure~2, top panel).
The blue box running from 22.2--23.6\AA\ marks the wavelength region
where these complications make reliable determination of residuals
difficult.  We have normalized the residuals of the two smoothed spectra
so that, in each case, the range $-1$ to $+1$ contains 68\% of
the data points between 21\AA\ and 27.7\AA, excluding this problematic region.
Dotted horizontal lines mark the symmetric range about zero that
contains 90\% (upper panel) or 95\% (lower panel) of the data points
in the same wavelength regions.  Solid vertical line segments mark
the expected positions of the $\lambda$21.602\AA\ line of \ovii (green)
and the $\lambda$18.969\AA\ line of \oviii (red) at the redshifts
of the \ovi systems found by TSJ and \cite{oegerle00}:
$z=0.26659$, 0.24531, 0.22637, 0.22497, 0.21326, and 0.12137.
For brevity, we will often refer to these redshifts simply as z1--z6,
in descending order (right to left in Figure~1).
Squares below the line segments in the
upper panel have an area proportional to the equivalent width of
the $\lambda$1032\AA\ \ovi line reported by TSJ or \cite{oegerle00}.
The wavelength calibration uncertainty $\Delta\lambda=0.025$\AA\ 
corresponds to two channels in these histogram plots.  
As discussed in \S\ref{sec:observations}, this uncertainty is not
symmetric, and we expect a line associated with a given \ovi system
to appear in the predicted channel or in one of the two channels
to its right (at higher $\lambda$).
In velocity units,
one channel is 0.0125\AA$\times (c/\lambda)/(1+z) \sim 125\kms$ 
(at $z=0.2$, $\lambda=25$\AA), and the wavelength calibration uncertainty
is $\sim 250\kms$.

Visual inspection of either panel of Figure~1 shows that three of
the deepest minima occur within the 2-channel wavelength calibration
uncertainty of the z2 and z6 \ovii wavelengths 
and the z6 \oviii wavelength 
(26.901\AA, 24.224\AA, and 21.271\AA, respectively).
Subsequent analysis (described below) shows the presence of absorption
features that are significant at the $\ga 2\sigma$ level at these
wavelengths, so we classify these systems as probable detections
and mark them with filled symbols in the upper panel of Figure~1.
The strongest \ovi system lies at z4=0.22497, and it has a weaker
neighbor at z3=0.22637.  There is a strong minimum in the residual
spectrum just redward of the predicted \ovii wavelengths,
but it is not consistent with
lying at either of the \ovi redshifts.  Since this line is 
comparable in significance to the \ovii lines at z2 and z6,
and since it lies within $1000\kms$ of the strongest \ovi system
(and within $650\kms$ of the weaker system at z3), we consider
it a candidate for an \ovii system at a nearby, but previously
unknown, absorption redshift, and we mark it by dotted green vertical
line segments in Figure~1.  In similar fashion, we consider the
feature $\sim 1000\kms$ redward of the apparent \oviii line
at z6 to be a candidate for a new \oviii absorption line, marked
by the dotted vertical red segments.  There are no strong signals
near the redshifts of the other \ovi systems.  The \oviii wavelength
of the z2 system, $\lambda=23.622$\AA, lies close to the Galactic \oi
absorption line, but we still would have been able to detect a
separate line if it were present and strong.
There is a strong feature at $\lambda \approx 25.35$\AA\ for which
we have no clear identification.  Its strength and breadth make
it unlikely to be intervening \oviin.

For quantitative assessment of features and measurement of line
parameters, it is preferable to work with the unsmoothed, directly
observed spectrum.  Figure~2 shows the 21-28\AA\ region of the
observed spectrum, with no extra smoothing or binning.
Starting with \oviin, we added to our continuum
model a Gaussian absorption line at the $\lambda$21.602\AA\ wavelength
for each of the \ovi redshifts z1 to z6.  The intrinsic FWHM of each line 
was fixed to be $100\kms$, similar to that of the \ovi lines (TSJ);
the precise value of the line width is unimportant, since the lines
are unresolved if they are narrower than $\sim 250\kms$.
The normalizations of the absorption lines were thus the only new
free parameters, which we determined by fitting the continuum plus
absorption lines model to the observed spectrum 
(using the continuum parameters determined
previously from the coarsely binned data),
folding in the effective area and the line response function.
We performed this and all subsequent fits using the
``sherpa''  software within CIAO-2.2.

This fit yielded non-zero normalizations of the absorption lines
at z2, z4, and z6, but not at z1, z3, or z5, just as one might
expect from visual inspection of Figure~1.
To allow for the wavelength calibration uncertainty and for the possibility
of small velocity offsets between \ovi and \oviin/\oviii absorption,
we refit the spectrum allowing the positions of the z2, z4, and z6
lines to vary.  This fit yielded offsets
of $\Delta\lambda=+0.0111$\AA\ and $+0.0274$\AA\ at z2 and z6,
within the wavelength calibration uncertainty, while the offset
at z4 was significantly larger.
We then repeated the model fit a final time, this time fixing
the z2 and z6 line positions to these best-fit values and the z4
line position to the original value based on the \ovi redshift,
with the normalizations of the z2, z4, and z6 systems as the 
free parameters.  Table~1 lists the corresponding equivalent widths
of the three fitted lines.  The reduction in $\chi^2$ obtained
by adding each of the three lines is $\Delta\chi^2=4.8$, 1.27,
and 2.68, with a corresponding $F$-test statistical significance
(for one additional degree of freedom in each case) of
99.28\%, 83.8\%, and 95.8\%, respectively.  We therefore classify
the z2 and z6 lines, but not the z4 line, as probable detections,
significant at the $2-3\sigma$ level.  The $1\sigma$ error bars
quoted in Table~1 correspond to $\Delta\chi^2=1$ relative to 
the model with best-fit parameters.

Applying the same procedure at the expected wavelengths of the
\oviii $\lambda$18.969\AA\ line yielded a significant detection
at z6, a non-zero value at z2, and best-fit normalizations of
zero at the other \ovi redshifts, just as expected from Figure~1.
For the z6 line, there is a wavelength offset of +0.0242\AA,
and the reduction in $\chi^2$ is 3.51, giving
an $F$-test significance of 98.03\%.  The z2 line reduces $\chi^2$
by 1.08 (significance 80.4\%).

We also searched for absorption in the same way at the expected
wavelengths of the 
\neix $\lambda 13.447$\AA\ and \nex $\lambda 12.1337$\AA\ lines. 
This analysis yielded a line at the \neix wavelength at redshift z4,
with $\Delta\chi^2=4.86$, and an $F-$test significance for
one additional degree of freedom of 98.3\%.
The existence of detectable \neix for this system,
which has at best weak \ovii and \oviii absorption, 
would be remarkable, implying non-solar
abundance ratios.  \cite{nicastro02} find evidence for super-solar
Ne/O at z$\sim 0$.  If the \neix detection here is real,
the implied Ne/O ratio for this system would be higher still.
There were no significant detections of \neix or \nex lines for
any of the other \ovi systems, nor of X-ray lines from
\civn, \cvn, \nvin, or \nviin.

We determined $1\sigma$ upper limits on the strength of the \oviin,
\oviiin, and \neix lines at the redshifts of the remaining \ovi
systems by adding lines at the expected wavelength and finding
the amplitude that produced $\Delta\chi^2=1$ relative to the
smooth continuum model.  These upper limits are listed in Table~1.
The \oviii limits for the z3, z4, and z5 systems are somewhat
less secure than the others because of the complicated form
of the continuum in this wavelength region.
We also fit line parameters for the two candidate ``new'' systems 
discussed earlier, one close to the z3-z4 \ovii wavelength
(with $z=0.229048$) and one close to the z6 \oviii wavelength
(with $z=0.125847$).  The derived equivalent widths and $1\sigma$
error bars for these systems are listed in Table~2.
Although these systems lie within $\sim 1000\kms$ of known \ovi
redshifts, this condition is much less stringent than that of
matching within the $0-250\kms$ wavelength calibration uncertainty,
making the probability of chance coincidence much higher.
We therefore attribute substantially less significance to
these two candidate systems, even though they are comparable in
strength to the ``detected'' \ovii and \oviii lines listed in Table~1.

Our final model for the 21--28\AA\ region of the spectrum,
therefore, consists of the continuum model described previously,
plus \ovii absorption lines at z2, z4, z6, and the
``new'' absorption redshift $z=0.229048$, plus \oviii absorption
lines at z2, z6, and the ``new'' redshift $z=0.125847$.
The z4 \ovii and z2 \oviii features are different from zero
at the $\sim 1\sigma$ level, while the other features are 
different from zero at the $\ga 2\sigma$ level.
Figure~2 shows the observed spectrum and this continuum+lines model,
over the full 21--28\AA\ region, and in close-up views near the
absorption redshifts.  Figure~3 shows the 14--18\AA\ region in
the same fashion.  Here the model consists of the continuum with
a single \neix absorption line at z4, which is different from
zero at the $2-2.5\sigma$ level.

Since the signals we are looking for are weak, and the detection
of even one X-ray forest system of great physical import,
we have been as meticulous as possible in our data reduction
and analysis.  In addition to the analysis reported here, 
we analyzed several data sets in the \chandra archive that 
have similar instrument set-up to search for any hidden systematic
effects, e.g., instrumental features that would produce artificially
low counts in the neighborhood of our detected lines.
We found no evidence for any such effects, and we are therefore
confident that the features seen in Figures~1--3 are true
properties of the data, not artifacts of the observational
or analysis procedures.

The existence of features with $\ga 2\sigma$ significance at several
wavelengths predicted {\it a priori} on the basis of \ovi absorption 
suggests that we
have indeed detected X-ray forest lines from highly ionized oxygen
in these systems.  However, we cannot rule out the possibility that
the coincidence between observed features and known absorption redshifts
is, in fact, just a coincidence.
It is difficult to estimate the probability of obtaining our results
``by chance'' if there were no true signals in the spectrum,
since the novelty of the observation was such that we could not
define clear ``detection'' criteria in advance of seeing the data.
Very roughly, we can note that 3 of the 12 strongest negative deviations
in the 21--28\AA\ region of the residual spectrum lie within the
0.025\AA\ wavelength calibration uncertainty of a known \ovi redshift.
(This is true for either of the residual spectra shown in Figure~1.)
Excluding the problematic 22.2--23.6\AA\ region, there are about
450 channels in this spectral range, and nine absorption wavelengths 
(six \ovii and three \oviiin) that could serve as possible ``matches''
to a given feature.
The wavelength calibration uncertainty gives any of the expected
absorption wavelengths a 3-channel range for a ``match,'' so
the probability that a randomly selected channel in this wavelength range
matches one expected absorption system is about $9\times 3/450 =0.06.$  
The probability of getting 3 of the 12 strongest deviations matched
with known redshifts, in the absence of a true physical correlation,
is then $(0.06)^3$ multiplied by $12!/(9!\times 3!)=220$ (the number
of ways that one can choose 3 distinct objects from a set of 12),
or about 5\%.  Since this probability is not extremely small,
we will be cautious in interpreting our
observations, considering both the possibility that we have true detections 
of several X-ray forest absorbers and the possibility that we have
only upper limits.  The only way to remove this ambiguity is to 
obtain a higher S/N spectrum with a longer observation.

\section{Implications}
\label{sec:implications}

The high density of weak \ovi absorbers establishes them as an
important constituent of the low redshift universe.
TSJ estimate $dN/dz \sim 48$ for \ovi absorbers with 
rest-frame equivalent width $W_r \ga 30\,$m\AA, based on
{\it HST}/STIS observations of the H1821+643 sightline.
While the statistical error on this estimate is large because of
the small number of systems, the discovery of an additional
low-$z$ \ovi system by \cite{oegerle00} and of a comparable line density
towards PG0953+415 by \cite{savage02} supports the inference of a high
cosmic incidence of \ovi absorption (see \citealt{savage02} for discussion).
By constraining the absorption produced
by more highly ionized oxygen species, our X-ray
data provide valuable new constraints on the physical state
of the \ovi absorbers.

As emphasized by \cite{hellsten98}, photoionization by the cosmic X-ray
background can have an important impact on the fractional abundance
of \ovii and \oviii at moderate overdensities, and photoionization
by the UV background can strongly influence the fractional abundance
of \ovin.  We therefore follow the methodology of \cite{chen02}
and calculate $f(\oviin)/f(\ovin)$ and $f(\oviiin)/f(\ovin)$ as
a function of density and temperature using the publicly available
code CLOUDY \citep{ferland99}, incorporating photoionization
by the UV background as estimated by \cite{shull99} and the 
soft X-ray background estimated by \cite{miyaji98}. 
Figure~4 shows the fractional abundance of \ovn, \ovin,
\oviin, and \oviii as a function of temperature for collisional
ionization and for collisional + photoionization
in gas with hydrogen number
density $n_H=10^{-6}\cm^{-3}$, $10^{-5}\cm^{-3}$, and $10^{-4}\cm^{-3}$.
For $\Omega_b h_{70}^2 = 0.04$, these physical densities correspond
to overdensities $\delta_b \equiv \rho_b/\bar{\rho_b} = 6/(1+z)^3$,
$60/(1+z)^3$, and $600/(1+z)^3$, respectively.  
For reference, the overdensity at the ``virial boundary'' of a collapsed
halo is $\delta \sim 50-150$ in a flat universe with a cosmological
constant and $\Omega_m\sim 0.3$, with the precise value depending on
the halo profile and on the adopted definition of the virial radius
(see, e.g., \citealt{navarro97}).
At low temperatures
and low densities, the fractional abundances are strongly
influenced by photoionization, while at higher temperatures and
densities they approach the abundances expected in collisional
equilibrium.

Figure~5 translates these results into contours in the
 $f(\oviiin)/f(\ovin)$ vs.\ $f(\oviin)/f(\ovin)$ plane.
In the left panel, curves show tracks at physical densities
that correspond to 
$\delta_b (1+z)^3 =1$, 10, $10^2$, and $10^3$, with an additional
track for collisional ionization.  
Numbers along the tracks mark temperatures log$\;T=4.5$, 5.0, 5.5, 6.0;
for $T \geq 10^{6.5}\K$, $f(\oviin)/f(\ovin) > 1000$ for any overdensity.
The right panel shows lines of constant temperature.
For $T \la 10^{5.2}\K$, the ion ratios follow a diagonal track in
this plane that is essentially independent of temperature.
As $T$ increases, the tracks become more vertical, separating
along the $f(\oviin)/f(\ovin)$ dimension.  Comparing the two panels
shows that $f(\oviin)/f(\ovin)$ is primarily a diagnostic of gas
temperature, while $f(\oviiin)/f(\ovin)$ constrains the gas density
for a given value of $f(\oviin)/f(\ovin)$.  This behavior reflects
the competing roles of photoionization and collisional ionization,
with the latter being more important for higher temperatures,
higher densities, and lower ionization states.

To estimate 
$f(\oviin)/f(\ovin)$
and 
$f(\oviiin)/f(\ovin)$
for the observed systems, we need to convert line equivalent
widths to corresponding column densities for each ion species.
The second column of Table~3 lists the \ovi column densities
and $1\sigma$ error bars that TSJ and \cite{oegerle00} derive
by fitting profiles of the 1032\AA\ line.  Our data do not have
the resolution and S/N required for profile fitting, so we instead
infer \ovii and \oviii column densities using the relations for
optically thin lines,
\begin{eqnarray}
N(\oviin) &=& 3.48 \times 10^{15}\cm^{-2} 
                  \left(\frac{W_{\rm thin}}{10\;\hbox{m\AA}}\right) , 
                  \label{eqn:ovii}\\
N(\oviiin) &=& 7.56 \times 10^{15}\cm^{-2} 
                  \left(\frac{W_{\rm thin}}{10\;\hbox{m\AA}}\right) , 
                  \label{eqn:oviii}
\end{eqnarray}
where $W_{\rm thin}$ is the rest-frame equivalent width (note that the
observed-frame equivalent widths in Tables~1 and~2 must be divided by
$1+z$ before applying these relations).  Equations~(\ref{eqn:ovii})
and~(\ref{eqn:oviii}) are based on oscillator strengths of 0.70 and
0.42 for \ovii K$\alpha$ and \oviii K$\alpha$, respectively, taken
from \cite{verner96}.  Columns~3 and~4 of Table~3 list the
corresponding column density estimates with $1\sigma$ error bars, or
$1\sigma$ upper limits for lines with a best-fit equivalent width of
zero.  The z2 system has an inferred \ovii column density of $3.9 \pm
1.7 \times 10^{15}\cm^{-2}$, while the z6 system has $N(\oviin)=2.8
\pm 1.5\times 10^{15}\cm^{-2}$ and $N(\oviiin)=6.7 \pm 3.5 \times
10^{15}\cm^{-2}$. These are consistent with the upper limits derived
by \cite{fang02b} from a short \chandra exposure.  Upper limits for
undetected systems are typically $\sim 1.9\times 10^{15}\cm^{-2}$ for
\ovii and $\sim 3\times 10^{15}\cm^{-2}$ for \oviii.  The simulations
of \cite{chen02} suggest that saturation effects for typical lines are
small ($\la 30\%$) at column densities $N(\oviin) \leq 4\times
10^{15}\cm^{-2}$ or $N(\oviiin) \leq 10^{16}\cm^{-2}$.  We therefore
expect saturation corrections to be unimportant in the case of our
upper limits and limited but not entirely negligible in the case of
our detections.  Figure~6 shows the curve-of-growth relation for \ovii
and \oviii lines with $b$ parameters in the range $50-250\kms$, from
which one can read off the column density as a function of
(rest-frame) equivalent width for an assumed value of $b$.

In Figure~5, points with $1\sigma$ error crosses mark the two detected
\ovi systems at z2 and z6, with ion ratios based on the numbers
in Table~3.  Since the fractional errors in \ovi column densities
are far smaller than those in the \ovii or \oviii column densities,
we have ignored them in calculating the errors on column density ratios.
For the remaining four systems, we plot arrows showing the $1\sigma$
upper limits on $f(\oviin)/f(\ovin)$ and $f(\oviiin)/f(\ovin)$.
Since the upper limits on the strength of the X-ray forest lines
are roughly constant from system to system, the upper limits on the
ion ratios are, to a first approximation, inversely proportional
to the \ovi column density.  The main exception is the \ovii line
of the z4 system, for which the best-fit column density is $1\sigma$
different from zero.  The best-fit $f(\oviin)/f(\ovin)$ ratio of
this system lies near the left-hand edge of the horizontal arrow.

At the $\sim 1\sigma$ level, our measurements and limits have a number
of interesting implications.  First, there are variations in the ion
ratios from system to system: the lower limit on $f(\oviin)/f(\ovin)$
for the z2 system is above the upper limit for the z4 and z1 systems,
and the lower limit on $f(\oviiin)/f(\ovin)$ for the z6 system is
above the upper limit for the z4 system.  Given the sensitivity of the
abundance ratios to density and temperature, it is not particularly
surprising to find that they vary from system to system, but these
results suggest that the diversity in physical properties will be
important in understanding the contribution of \ovi systems to the
baryon budget.  The z2 and z6 systems have higher
$N(\ovin)/N(\hbox{H{\sc i}})$ than the other four systems (TSJ;
\citealt{oegerle00}), perhaps implying higher gas temperatures that
also account for the higher relative abundance of \ovii and \oviiin.
Second, the upper limits on the undetected systems restrict their
locations in the temperature-density plane --- roughly speaking, the
absence of detectable \ovii provides an upper limit on the gas
temperature, and the absence of detectable \oviii provides a lower
limit on the gas density for a given temperature.  For example, the z4
system must have $T \la 10^{5.5}\K$, and if it is close to this
temperature, it must have $\delta_b \ga 5$.  Third, the overdensities
implied by the best-fit line parameters of the detected systems at z2
and z6 are significantly below the values $\delta_b \sim 100$
corresponding to virialized systems.  In physical terms, the
co-existence of detectable amounts of \ovin, \oviin, and \oviii
requires that photoionization play a central role in determining the
abundance ratios, which is possible only if the density is fairly low.

The last of these conclusions is the most interesting, and also the
most surprising.  In the case of the z2 absorber, the density implied
by the best-fit abundance ratios is $n_H \sim 10^{-5}\cm^{-3}$,
corresponding to $\delta_b(1+z)^3 \sim 60$.  If we assume that
$f(\oviin)+f(\oviiin) \approx 1$, then the path length required to
produce the estimated column density $N(\oviin)+N(\oviiin) = 7.1\times
10^{15}\cm^{-2}$ is $L = 3\;{\rm Mpc}\;(Z/0.1Z_\odot)^{-1}$, where $Z$
is the metallicity and we take solar relative abundances, $n_O/n_H =
7.41 \times 10^{-5} (Z/0.1Z_\odot)$, from \cite{verner96}.  This path
is rather long for a system of $T \sim 10^{5.8}\K$, unless the
metallicity is significantly above $0.1Z_\odot$, but the ion ratios of
this system are consistent at the $1\sigma$ level with a density an
order-of-magnitude higher, which would reduce the implied path length
by the same factor.  The z6 system is more of a puzzle.  Taking
$\delta_b(1+z)^3 \sim 10$ (higher than the best-fit value, but easily
within the $1\sigma$ error bar), the path length implied by a similar
calculation is $L\sim 24\;{\rm Mpc}\;(Z/0.1Z_\odot)^{-1}$, which would
be long even for $Z=Z_\odot$.

One possibility, of course, is that the true abundance ratios of this
system lie outside the range of our $1\sigma$ error estimates.  In
particular, if $f(\oviiin)/f(\ovin)$ is substantially lower than our
estimate, then the density of the system could be much higher, and the
implied path length shorter.  The implied density would also be higher
if the inferred \ovii column density has been artificially depressed
by saturation.  A third possibility is that the absorber is
multi-phase, with the \oviii (and perhaps \oviin) absorption arising
in hotter gas.  In this case, the ion ratios do not yield accurate
constraints on the density.  Higher S/N observations could rule out
(or confirm) the first possibility, and good constraints on the line
width from UV observations could address the second, but the last is
an unavoidable source of systematic uncertainty in the interpretation
of measurements like these.

Unfortunately, our constraints at the $2\sigma$ level are rather weak.
The \ovii and \oviii fractions of the detected systems are consistent
with zero at the $\sim 2\sigma$ level.  One can see the approximate
$2\sigma$ upper limits for these systems in Figure~5 by adding
log$\;3/2 \approx 0.18$ to the points.  Similarly, one obtains
approximate $2\sigma$ upper limits for the undetected systems by
adding log$\;2\approx 0.3$ to the $1\sigma$ upper limits shown in
Figure~5.  The conclusion that $T < 10^6\K$, with a tighter upper
limit for stronger absorbers, holds robustly, but stronger physical
constraints on the \ovi systems at the $2\sigma$ level require higher
S/N than our present observations afford.

What can we say about the contribution of \ovi absorbers to the cosmic
baryon budget?  TSJ estimate this contribution based on their STIS
observations of H1821+643, using the equation
\begin{equation}
\Omega_b(\ovin) = {\mu m_H H_0 \over \rho_c c} {1 \over \Delta X}
  \left(\frac{\rm O}{\rm H}\right)^{-1}
  \left\langle 
    \left[f(\ovin)\right]^{-1} \right\rangle 
  \sum_i N_i(\ovin) ~.
\label{eqn:omb1}
\end{equation}
Here $\rho_c$ is the critical density, $\Delta X$ is the absorption 
distance interval \citep{bahcall69} probed by the observations,
%$({\rm O}/{\rm H})$ is the ratio of oxygen to hydrogen by number,
and $N_i(\ovin)$ is the \ovi column density in system $i$.
The factor $({\rm O}/{\rm H})^{-1}[f(\ovin)]^{-1}$ converts the \ovi
column density of a system to the hydrogen column density, and
the average represented by $\langle ... \rangle$ should be weighted
by \ovi column density for equation~(\ref{eqn:omb1}) to hold.
In fact, $({\rm O}/{\rm H})^{-1}$ should be inside the $\langle ... \rangle$
as well, but since our observations give no additional purchase
on the metallicity, we will assume that it is constant from 
system to system.
Adopting uniform values of $[{\rm O/H}]=-1$ and $f(\ovin)=0.2$, TSJ
find $\Omega_b(\ovin)=0.0043 h_{70}^{-1}$.  Since $f(\ovin)=0.2$
is close to the maximum possible \ovi fraction, TSJ interpret this
estimate as a lower limit on $\Omega_b(\ovin)$ for $[{\rm O/H}]=-1$.
More generally, we can express the TSJ result as
\begin{equation}
\Omega_b(\ovin)_{\rm TSJ} = 0.0043 \; h_{70}^{-1} \; 
                \left(\frac{0.1}{10^{[{\rm O/H}]}}\right) \;
                \left(\frac{\langle [f(\ovin)]^{-1} \rangle}{5.0} \right)~.
\label{eqn:tsj}
\end{equation}
Subsequent observations of three other sight lines
\citep{richter01,sembach01,savage02} suggest that the density of
\ovi absorbers towards H1821+643 may be higher than average, 
and that the TSJ estimate should perhaps
be revised downwards by about a factor of two \citep{savage02}.

While small number statistics and the unknown metallicity are
both significant sources of uncertainty in the $\Omega_b(\ovin)$
estimates, the major uncertainty in equation~(\ref{eqn:omb1})
is the \ovi ionization fraction, since 
$\langle [f(\ovin)]^{-1} \rangle$
could quite plausibly be an order-of-magnitude larger than
the factor of 5.0 adopted by TSJ.
However, since $f(\ovin)+f(\oviin)+f(\oviiin)$ should be close to unity
in most systems that have detectable \ovi (see Figure~4), it is just
this factor that the \chandra spectrum allows us to estimate:
\begin{equation}
\langle [f(\ovin)]^{-1} \rangle \approx 
  {\sum_i N_i(\ovin) + N_i (\oviin) + N_i(\oviiin) \over
              \sum_i N_i(\ovin)} ~.
\label{eqn:fo6}
\end{equation}
The sums in equation~(\ref{eqn:fo6}) weight $[f(\ovin)]^{-1}$ by
\ovi column density, as desired for the $\Omega_b(\ovin)$ estimate.

For our {\it estimate} of $\langle [f(\ovin)]^{-1} \rangle$, we use
the measured values of $N(\oviin)$ and $N(\oviiin)$ for the z6 system
and of $N(\oviin)$ for the z2 system; we use $N(\oviiin)=0$ for the z2
system, and we set $N(\ovin)+N(\oviin)+N(\oviiin)=5N(\ovin)$ for the
other four systems.  In other words, we treat all of our $2\sigma$
detections as real, but where we have only $1\sigma$ measurements or
upper limits, we conservatively assume the lowest \ovii and \oviii
column densities consistent with the physical expectation that
$f(\ovin)\leq 0.2$.  To obtain a $1\sigma$ error bar on this estimate,
we sum the errors of the three detected X-ray lines in quadrature, and
we ignore the much smaller contribution from uncertainty in the \ovi
column densities.  This calculation yields $\langle [f(\ovin)]^{-1}
\rangle = 32 \pm 9$.  Combined with the TSJ result
(eq.~\ref{eqn:tsj}), this ratio implies $\Omega_b(\ovin) = 0.028 \pm
0.008 \; h_{70}^{-1}$ for $[{\rm O/H}]=-1$.

The statistical error bar here accounts {\it only} for the
observational uncertainties in the $N(\oviin)$ and $N(\oviiin)$
measurements, not the uncertainty due to the small number of systems
along the single line of sight employed in the analysis, which is
probably at least as important.  Furthermore, each of our detected
features is consistent with zero at the $\sim 2\sigma$ level, and we
cannot rule out the possibility that the apparent association between
these features and \ovi redshifts is coincidental.  Nonetheless, this
analysis of the \chandra spectrum of H1821+643 provides the first
direct evidence (a) that the column density-weighted mean value of
$[f(\ovin)]^{-1}$, is substantially higher than the conservative value
of 5.0 that TSJ and \cite{savage02} used to derive lower limits on
$\Omega_b(\ovin)$, and (b) that as a result, the baryon fraction
associated with \ovi absorbers substantially exceeds that of any other
known low redshift baryon component, representing an appreciable
fraction of the baryons predicted by BBN.  The combined column density
of the candidate ``new'' \ovii and \oviii lines is comparable to that
of the three lines at \ovi redshifts, so these systems, if real, would
represent a hotter IGM component containing a similar baryon fraction.

The estimate of $\langle [f(\ovin)]^{-1} \rangle$ requires that we
treat the $2\sigma$ lines as real detections.  More conservatively, we
can derive an upper limit on $\langle [f(\ovin)]^{-1} \rangle$,
without taking a stand one way or the other on the reality of the
detections.  Of course, since our conservative estimate already
implies $\Omega_b(\ovin)$ of the same order as $\Omega_{\rm BBN}$,
these upper limits are not very restrictive.  For the $1\sigma$ upper
limit, we carry out the sum in equation~(\ref{eqn:fo6}) including
$N(\oviiin)$ from the z2 system and $N(\oviin)$ from the z4 system,
and we add to it the quadrature sum of the $1\sigma$ errors on
$N(\oviin)$ and $N(\oviiin)$ for all six systems, dividing the total
by $\sum_i N_i(\ovin)$.  For the corresponding $2\sigma$ upper limit,
we simply double the error bars on each system.  The result is
$\langle [f(\ovin)]^{-1} \rangle < 60$ at $1\sigma$ and $\langle
[f(\ovin)]^{-1} \rangle < 79$ at $2\sigma$.  In combination with the
TSJ or \cite{savage02} numbers, even the $1\sigma$ upper limit is
consistent with $\Omega_b(\ovin) \approx \Omega_{\rm BBN}$.  If we had
obtained null results for all of the \ovi systems, then the upper
limit on $\Omega_b(\ovin)$ would have come out well below $\Omega_{\rm
BBN}$.

\ovii lines with observed-frame $W \ga 26$m\AA, or \oviii lines with
$W \ga 20$m\AA, would have appeared in our data unambiguously, with
significance $\ga 4\sigma$.  We can clearly rule out the existence of
such systems along this line of sight.  If we exclude the
$22.2-23.6$\AA\ region, then the observable path length for the \ovii
systems is $\Delta z=0.2322$ and for \oviii systems $\Delta z=0.2232$.
We therefore estimate that the number density of such systems is
$dN/dz < (\Delta z)^{-1}\sim 4$ in each case.  For Poisson statistics,
the probability of finding no systems where three are expected is
$e^{-3}\approx 0.05$, so the 95\% confidence limit on the density of
strong \ovii and \oviii lines is $dN/dz \la 12$.

In fact, even the existence of the systems that we have detected at
the $2-3\sigma$ level is rather surprising relative to theoretical
expectations.  For IGM metallicity $[{\rm O/H}]=-1$, hydrodynamic
simulations predict one \ovii system per unit redshift above column
density $N\sim 10^{15}\cm^{-2}$, with a similar result for \oviiin,
and a rapid decline in the line density for higher column density
thresholds.  We base this statement on Figure~7 of \cite{chen02}, but
the comparisons in that paper suggest that the \cite{hellsten98} or
\cite{fang02a} simulations would yield similar results for the same
metallicity assumption, at least within a factor $\sim 2$.  Adding
scatter to the IGM metallicities increases the probability of finding
strong \ovii or \oviii systems, since there are more low (total)
column density systems to scatter to high (oxygen) column densities
than {\it vice versa}.  With the metallicity scatter predicted by
\cite{cen99b}, the line density predicted by \cite{chen02} falls to
$dN/dz\sim 1$ at thresholds of $N(\oviin) \sim 2\times
10^{15}\cm^{-2}$ and $N(\oviiin)\sim 3\times 10^{15}\cm^{-2}$.  It
remains surprising to find $2-4$ systems (two if we count only those
at \ovi redshifts, four if we count the candidate ``new'' systems)
with column density in excess of $4\times 10^{15}\cm^{-2}$ in a path
length of $\Delta z \sim 0.23$.  The incidence of \ovi absorption
towards H1821+643 appears to be higher than average \citep{savage02},
and the same could be true for \ovii and \oviii absorption.
Nonetheless, accounting for the systems reported here within the
framework of the $\Lambda$CDM cosmological model (or at least the Chen
et al. [\citeyear{chen02}] simulation of it) requires that some
regions of the shock-heated IGM be enriched to metallicities several
times higher than $[{\rm O/H}]=-1$, and it probably requires that the
frequency of such enriched regions be higher than the \cite{cen99b}
model predicts.  Confirmation of the observed systems at higher S/N,
and correspondingly improved estimates of the \ovii and \oviii column
densities, would allow firmer assessment of the theoretical
predictions for X-ray forest absorption.

\section{Conclusions and Outlook}
\label{sec:conclusions}

In combination with the {\it HST} and {\it FUSE} observations of \ovi
absorption, this {\it Chandra} spectrum of H1821+643 provides the most
direct evidence to date for the pervasive, moderate density,
shock-heated IGM predicted by leading cosmological models.  The
spectrum shows absorption signals with $\ga 2\sigma$ significance at
the expected wavelengths of \ovii and \oviii absorption for one of the
\ovi systems, at the expected wavelength of \ovii absorption for
another, and at the expected wavelength of \neix absorption for a
third.  There are two absorption features of comparable strength
within $1000\kms$ of predicted \ovii and \oviii absorption
wavelengths, which could represent additional absorbing gas in the
large scale environment of the \ovi systems.  Some of the other $\sim
2\sigma$ features in the spectrum could correspond to X-ray forest
systems without associated \ovi absorption, but these features are
present in roughly the number expected for Gaussian noise, so without
an {\it a priori} reason to search for absorption at their observed
wavelengths, we have no convincing evidence that they represent
physical systems.  There is a clear detection of Galactic \oi
absorption, and a strong, broad feature at 25.35\AA\, for which we
have no obvious identification.

The best estimates of the ratios $N(\oviin)/N(\ovin)$ and
$N(\oviiin)/N(\ovin)$ for the two detected \ovi systems imply that
they are largely photoionized and that gas densities are significantly
below those expected in virialized structures like groups or clusters.
This inference relies on the assumption that the \ovin, \oviin, and
\oviii absorption arise in the same gas, not in different components
of a multi-phase system.  Independent of the single-phase assumption,
the measured ratios imply that the column density-weighted mean value
of $[f(\ovin)]^{-1}$ is substantially higher than the value of 5.0
that TSJ used to derive their lower limits on $\Omega_b(\ovin)$, the
baryon density associated with \ovi absorbers.  TSJ and
\cite{savage02} find $\Omega_b(\ovin) \sim 0.002-0.004h_{70}^{-1}$ for
$[{\rm O/H}]=-1$ and $[f(\ovin)]^{-1}=5$, and our results imply that
$\langle [f(\ovin)]^{-1} \rangle = 32 \pm 9$ at $1\sigma$.  At face
value, therefore, our measurements suggest that the baryon fraction
associated with \ovi absorbers is substantially larger than that in
stars, cold gas, or hot gas detected in X-ray emission, and that the
\ovi absorbers trace a significant reservoir of the ``missing'' low
redshift baryons.  The candidate \ovii and \oviii systems that are not
at the \ovi redshifts contain a similar total amount of highly ionized
oxygen.  In combination with the results of hydrodynamic simulations
\citep{hellsten98,fang02b,chen02}, the existence of even one or two
X-ray forest lines at the column densities of our detected systems
suggests that some regions of the shock-heated IGM have been enriched
to well above $[{\rm O/H}]=-1$, and the oxygen and hydrogen column
densities measured for the z2 \ovi absorber suggest $[{\rm O/H}]\ga
-0.3$ in this system.

Unfortunately, our results also show that definitive measurements of
X-ray forest absorption are extraordinarily difficult, even with the
great technological advance that {\it Chandra}, LETG, and ACIS-S
represent over previous instruments.  The absorption signals for the
detected \ovi systems differ from zero at only the $2-3\sigma$ level,
and since we detect some but not all of the known \ovi systems, we
cannot rule out the possibility that the apparent association between
these features and predicted absorption redshifts is simply a
coincidence.  The features at other wavelengths are not strong enough
to represent clear detections of new systems.  The constraints on the
physical properties and associated baryon fraction of the \ovi
absorbers are very interesting at the $1\sigma$ level, but with
$2\sigma$ error bars we obtain only loose constraints that are not
particularly surprising, though they are still more than was
previously known.  {\it XMM-Newton} has larger effective area than
{\it Chandra} at soft X-ray wavelengths.  However, the effective area
near 25\AA\ is non-uniform, with many instrumental features at
observationally interesting wavelengths.  Furthermore, published
spectra obtained with RGS at these wavelengths seem to achieve lower
resolution than {\it Chandra}/LETG, reducing the sensitivity to narrow
lines.  It is not clear, therefore, whether {\it XMM-Newton} will
prove as powerful as \chandra in X-ray forest searches.

Studies of X-ray forest absorption are among the major scientific
drivers for 
{\it Constellation-X},\footnote{http://constellation.gsfc.nasa.gov} 
and for the still more powerful {\it XEUS} 
mission.\footnote{http://astro.estec.esa.nl/SA-general/Projects/XEUS}
{\it Chandra} and {\it XMM-Newton} can at best give us a glimpse
of what these future missions may achieve.
However, the tantalizing but still ambiguous hints of X-ray forest 
absorption in the H1821+643 spectrum would be frustrating to live with
for half a decade or more.  In our view, the
best step forward in the study of the X-ray forest would be
further {\it Chandra} observations of H1821+643, which remains
the best target for such an investigation because of its X-ray brightness
and well studied \ovi absorption, and because the {\it Chandra}
spectrum presented here provides 500 ksec of existing data and
clear objectives for a future observation.  
For example, an additional 1 Msec observation would reduce the
noise in the co-added spectrum by $3^{1/2}$, so any true physical
features that have $2\sigma$ significance in the current spectrum would
rise to $3.5\sigma$ significance, making their reality entirely
unambiguous, regardless of whether they lie at the redshift of a
known \ovi system.  Conversely, since we obtain $1\sigma$ upper
limits $N(\oviin) \la 1.9\times 10^{15}\cm^{-2}$ and
$N(\oviiin) \la 3\times 10^{15}\cm^{-2}$ for systems with no
absorption signal in the current spectrum, the $1\sigma$ upper
limits in undetected systems from a co-added 1.5 Msec spectrum would imply
$f(\oviin)/f(\ovin) < 11\; [N(\ovin)/10^{14}\cm^{-2}]^{-1}$ and
$f(\oviiin)/f(\ovin) < 18\; [N(\ovin)/10^{14}\cm^{-2}]^{-1}$, respectively.
These constraints are still not extremely strong for any individual
system, but null detections for all of the known \ovi systems in
such a spectrum would suggest that the total amount of material
they contain is not enough to represent the main reservoir
of the missing low redshift baryons.

{\it Chandra} observations can also guide the design of X-ray forest
studies with future missions.  If the apparent detections in our
current spectrum represent real systems, then {\it Constellation-X} is
likely to find an abundance of X-ray absorption lines tracing the
enriched, shock-heated IGM.  Conversely, if the evidence for these
features weakens as the S/N improves, then tall trees in the X-ray
forest are rare, and even {\it Constellation-X} will have to search
long and hard to find them.

\acknowledgments

We thank the entire \chandra team for a superb mission. We are
grateful to CXC scientists for extensive help in the data reduction
and analysis process.  We thank Jordi Miralda-Escud\'{e}, Fabrizio
Nicastro and Todd Tripp for helpful discussions.  This work is
supported by Chandra X-ray Observatory Grant GO1-2118X from
Smithsonian Astrophysical Observatory. DHW acknowledges the
hospitality of the Institut d'Astrophysique de Paris and support from
the French CNRS during the completion of this work. X.C. is supported
at Ohio State by the DOE under grant DE-FG03-92ER40701, and at
ITP/UCSB by the NSF under grant PHY99-07949.

\clearpage
\thispagestyle{empty}

%% If you use the table environment, please indicate horizontal rules using
%% Do not put multiple tabular environments within a single table.
%% The optional \label should appear inside the \caption command.

%% The following command ends your manuscript. LaTeX will ignore any text

\small
\begin{table}[h]
\small
\caption{Observed EWs of X-ray lines at the redshifts of OVI Absorbers$^1$}
\begin{tabular}{|lcccccc|}
\tableline
X-ray Lines & z1 & z2 & z3 & z4 & z5 & z6$^2$ \\
& & & & & & \\
\tableline
OVII K$\alpha$ & $<6.9$ & {\bf 13.9$\pm 6.2$} &$<6.3$ & 7.3$\pm 6.3$ & $<6.8$ & {\bf 9.1$\pm 4.7$} \\
& & & & & & \\
OVIII K$\alpha$ & $<4.7$ & 5.2$\pm 4.9$ & $<4.9$  & $<4.9$  & $<4.9$ & {\bf 9.9$\pm 5.2$}\\
& & & & & & \\
NeIX K$\alpha$ & $<3.7$ & $<3.7$ & $<3.6$ & {\bf 8.4$\pm 3.6$}  & $<3.6$ & $<3.4$ \\
& & & & & & \\
\tableline
\end{tabular}
\\
%\bigskip
1. All EWs are in m\AA, and errors are 1$\sigma$. Upper limits
(1$\sigma$) are listed when best fit values of EWs are
zero. ``Detected'' systems, with significance $\ga 2\sigma,$ are 
indicated in boldface type.  All EWs are {\it observed frame}
and must be divided by $(1+z)$ to obtain rest-frame values.
\\ 2. The redshifts are:
z1=0.26659, z2=0.24531, z3=0.22637, z4=0.22497, z5=0.21326 ,
z6=0.12137. Reference: z1--z5: Tripp et al. (HST); z6: Oegerle et al.
(FUSE) \\
\end{table}

\small
\begin{table}[h]
\small
\caption{Observed EWs of ``new'' systems$^1$}
\begin{tabular}{|lcc|}
\tableline
Redshifts & OVII K$\alpha$ & OVIII K$\alpha$$^2$ \\
\tableline
 & &  \\
z=0.229048 & 15.1$\pm 5.8$ & \nodata \\
 & &  \\
z=0.125847 & $<5.3$ & 11.1$\pm 5.0$  \\
 & &  \\
\tableline
\end{tabular}
\\
%\bigskip
1. Both new redshifts are within $\sim 1000$ km s$^{-1}$ of known
   OVI redshifts. \\
2. The OVIII K$\alpha$ line in the z=0.229048 system falls in the \oi edge.\\
\end{table}
\pagebreak

\newpage
\small
\begin{table}[h]
\small
\caption{Oxygen column densities in the intervening aborption systems$^1$}
\begin{tabular}{|lccc|}
\tableline
Redshifts & \ovi & \ovii & \oviii \\
\tableline
0.26659 & $5.1\pm 0.8$ & $<1.9$ & $<2.8$ \\
& & & \\
0.24531 & $5.2\pm 0.6$ & $3.9\pm1.7$ & $3.2\pm 3.0$ \\
& & & \\
0.22637 & $2.4\pm 0.5$ & $< 1.8$ & $< 3.0$\\
& & & \\
0.22497 & $19.9\pm 1.2$ & $2.1\pm 1.8$ & $< 3.0$\\
& & & \\
0.21326 & $3.55\pm 0.81$ &$< 1.9$ & $< 3.0$\\
& & & \\
0.12137 & $10.0\pm 2.0$ & $2.8\pm 1.5$ & $6.7\pm 3.5$ \\
& & & \\
\tableline
0.229048 & \nodata & $4.3\pm 1.6$ &  \nodata \\
& & & \\
0.125847 & \nodata & $<1.6$ & $7.5\pm 3.3$ \\
& & & \\
\tableline
\end{tabular}
\\
%\bigskip
\normalsize 1. The known \ovi systems are in the upper part of the table
and the ``new'' candidate systems in the lower part. 
\ovi column densities are in units of $10^{13}$ cm$^{-2}$. 
They are taken from TSJ and \cite{oegerle00} based
on the stronger (1032\AA) line of the doublet. 
\ovii and \oviii column densities are in units
of $10^{15}$ cm$^{-2}$.  They are derived from the equivalent
widths in Tables~1 and~2 assuming that the lines are optically
thin (eqs.~\ref{eqn:ovii} and~\ref{eqn:oviii}).
\end{table}

\newpage
\begin{figure} [h]
\centerline{
\epsfxsize=6.5truein
\epsfbox[35 330 550 700]{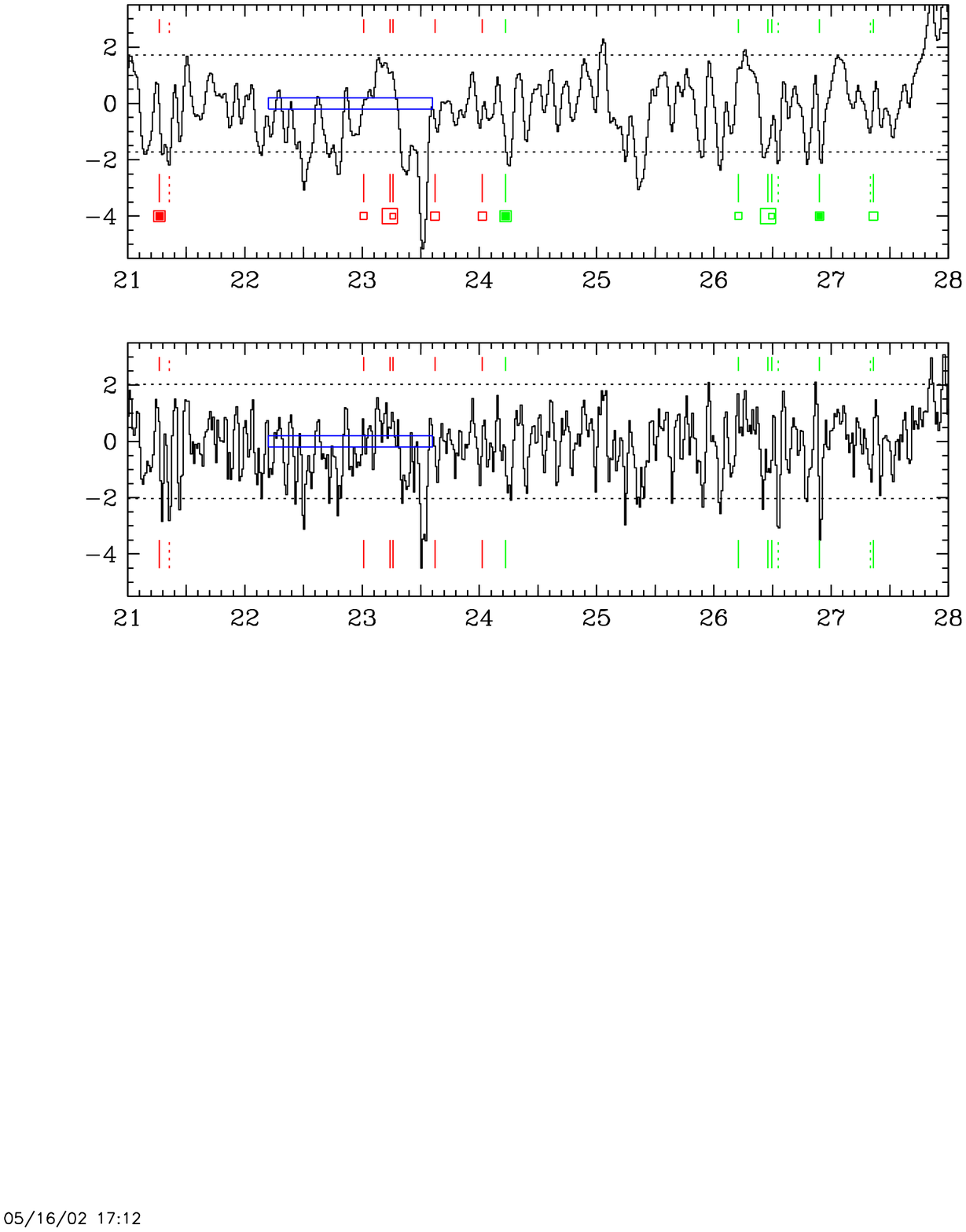}
}
\caption{
Normalized residual spectrum in the 21--28\AA\ region, where \ovii or \oviii
absorption associated with the known \ovi systems towards H1821+643
would lie.  The upper panel shows the residual spectrum convolved with
the line response function (LRF), a beta function of FWHM=0.05\AA\ 
(individual channels in the histogram are 0.0125\AA).
The lower panel shows the residual
spectrum convolved instead with a Savitsky-Golay filter, which provides
less noise suppression but better preserves the width and depth of
spectral features.  The blue box in each panel marks a wavelength
range (22.2--23.6\AA) in which Galactic absorption and instrument
response features make accurate continuum modeling difficult; the
strong absorption feature at 23.5\AA\ is Galactic \oin.  In each panel,
the residuals are multiplied by a normalization factor such that
the range $-1$ to $+1$ contains 68\% of the data values.
Dotted lines mark the symmetric intervals containing 90\% of the
data values in the upper panel and 95\% in the lower panel.
Solid line segments mark the expected positions of \ovii (green)
and \oviii (red) absorption at the known \ovi redshifts, z1 to z6
from right to left.
Squares in the upper panel are proportional in area to the
measured \ovi equivalent width.  Filled squares indicate ``detected''
systems, for which subsequent analysis shows a line at $\ga 2\sigma$
significance within the 0.025\AA\ (2-channel) wavelength calibration
uncertainty of the expected position.  Dotted line segments mark
other features of similar strength close to expected \ovii or \oviii
wavelengths.
}
\end{figure}

\newpage
\begin{figure} [h]
\psfig{file=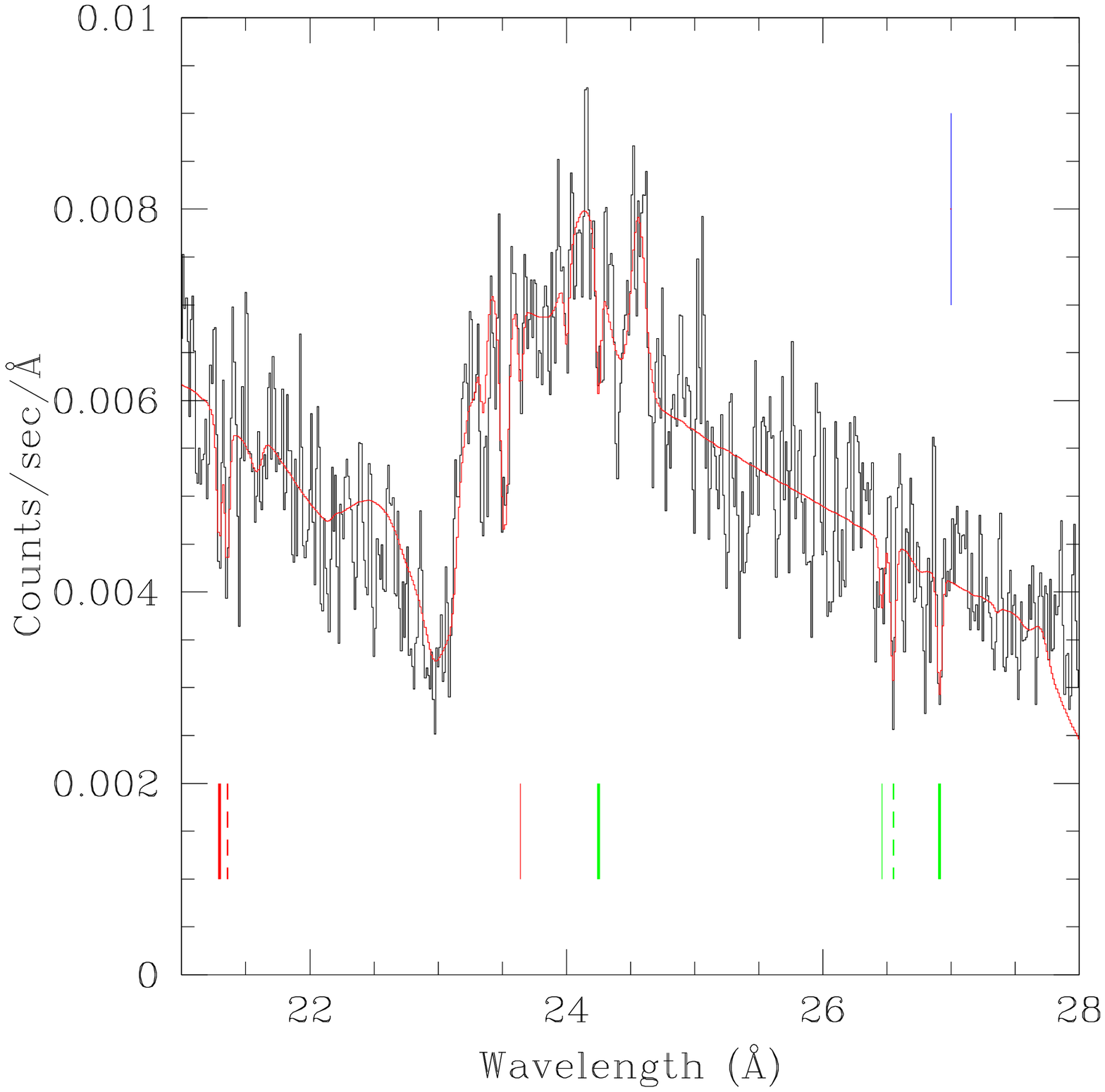,height=2.in,width=7.0in}
%\vspace*{-0.1in}
%\hspace*{1in}
\psfig{figure=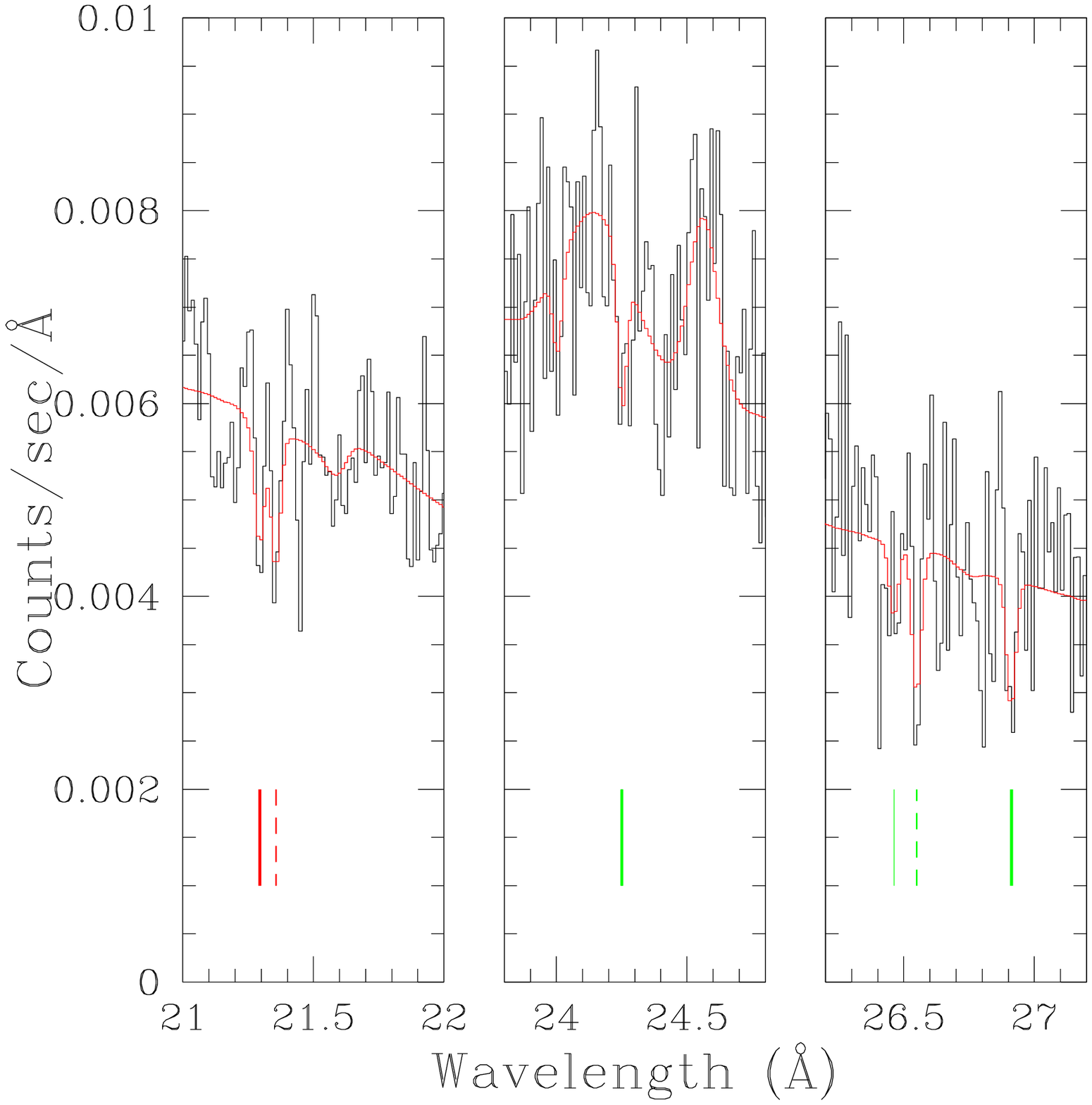,height=2.2in,width=7.0in}
\caption{Top: The $21-28$\AA\ region of the observed spectrum, with no
extra binning or smoothing.  The red line running through the spectrum
delineates the continuum $+$ absorption line model fit to the data.
The \ovii lines are marked in green and \oviii lines in red. The solid
tickmarks indicate lines at the known \ovi redshifts, with thick lines
indicating ``detected'' systems, whose significance is $\ga 2\sigma$.
The dashed tickmarks indicate the candidate ``new'' systems, which are
features of comparable strength within $1000\;\kms$ of known \ovi
redshifts.  The broad \oi absorption line at $\sim 23.5$\AA\ is from
the Galaxy and the O-K edge at $\sim 23$\AA\ is from the instrument as
well as the Galaxy. The blue bar in the upper right corner of the
figure represents a typical error-bar.  Bottom: Zoom on \oviii at z6
(left), \ovii at z6 (middle) and \ovii at z2 (right) regions showing
the detected absorption lines more clearly.  }
\end{figure}

\newpage
\begin{figure}[ht]
\psfig{figure=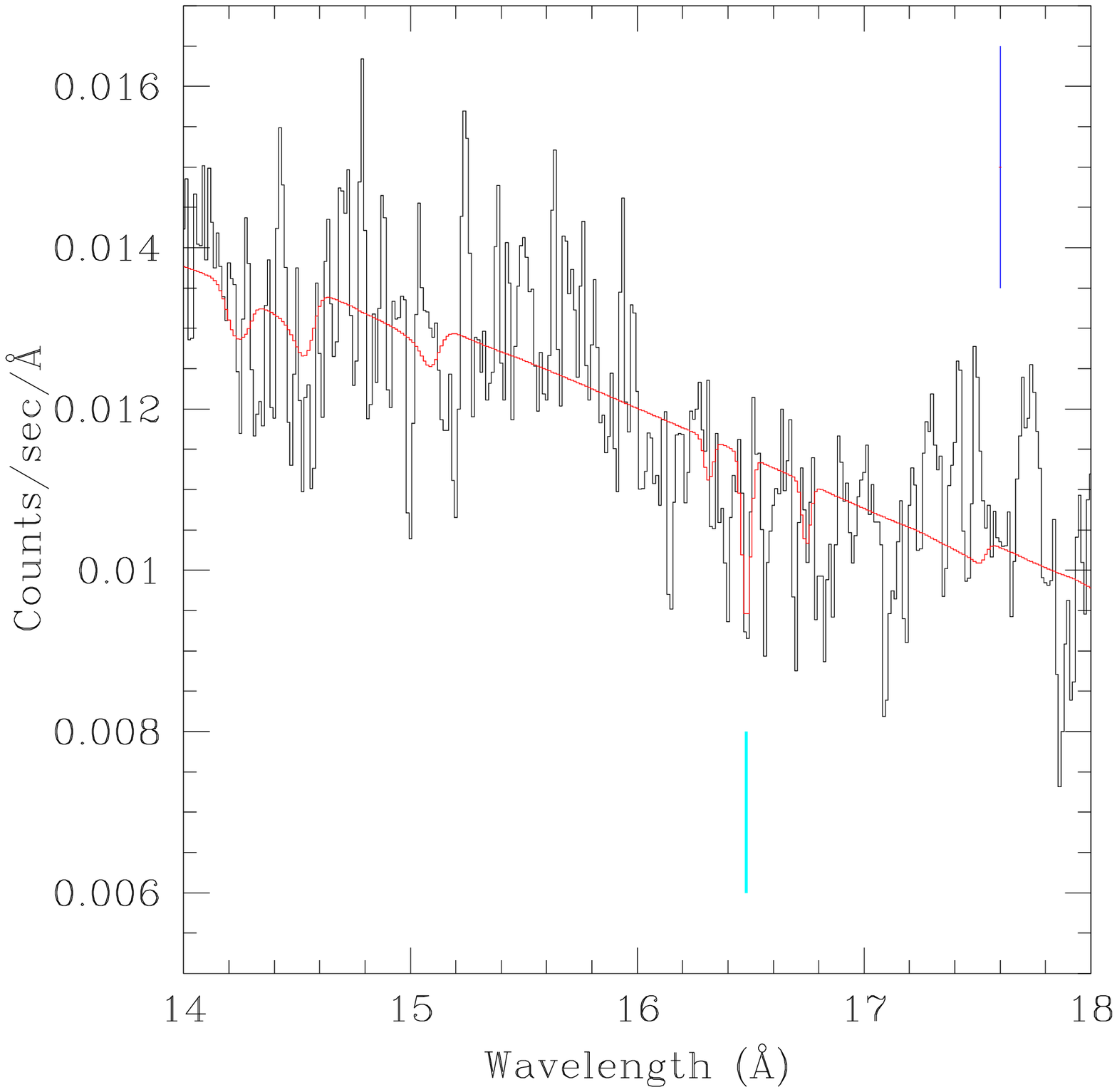,height=2in,width=7in}
%\vspace*{-0.1in}
\hspace*{2.5in}
\psfig{figure=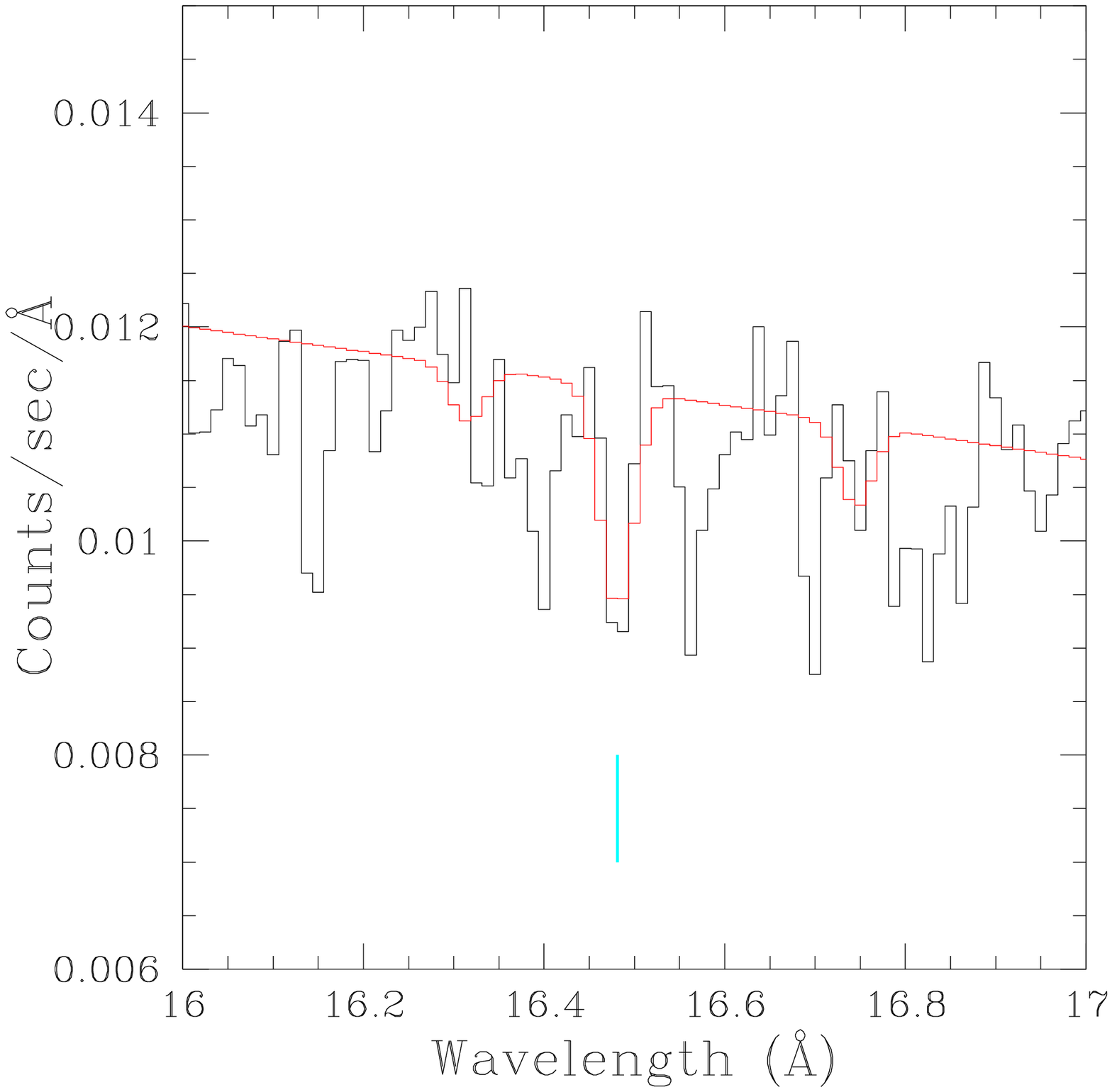,height=2.2in,width=2in}
\hspace*{1in}
\caption{Same as Figure 2, but in the NeIX region. The cyan tickmark
indicates the ``detected'' z4 system. Bottom: zoom on the z4 system.}
\end{figure}

%\newpage
%\begin{figure}[ht]
%\psfig{figure=fig4.ps,height=9.in,width=4.0in,angle=90}
%\caption{Best fit model of absorption lines to the unbinned data.}
%\end{figure}

\newpage
\begin{figure}[ht]
\centerline{
\epsfxsize=5.0truein
\epsfbox[30 160 565 690]{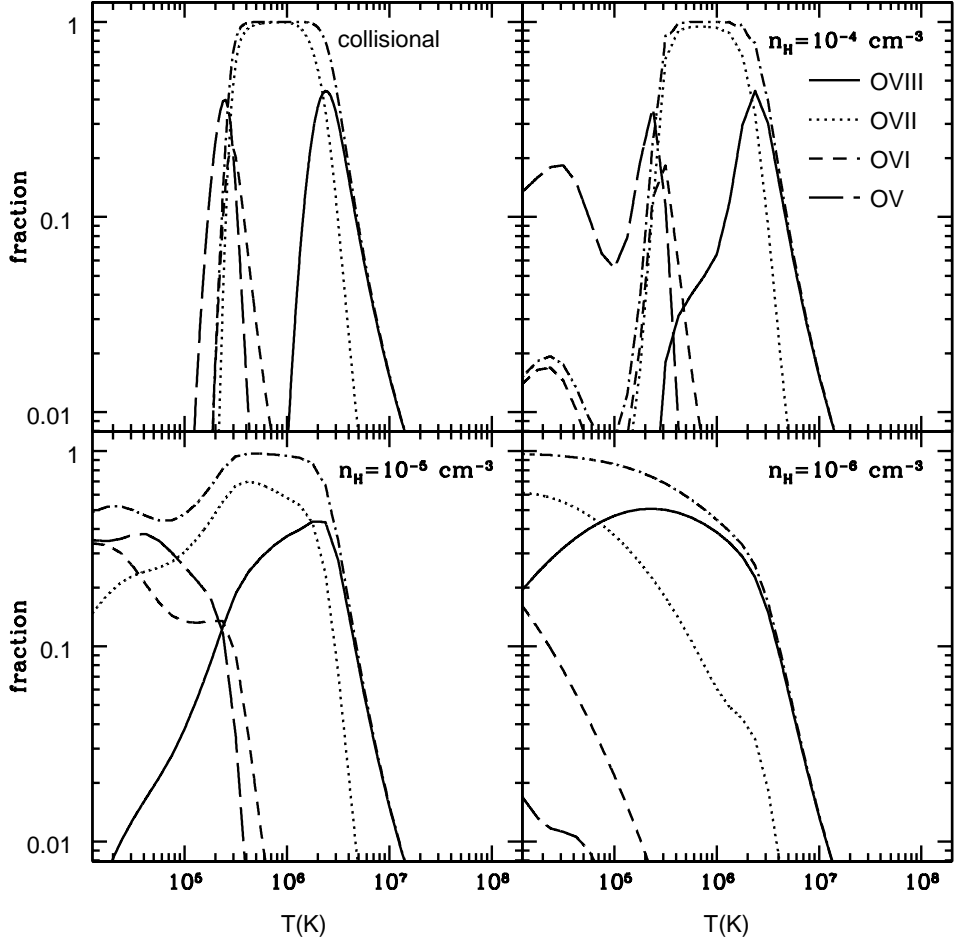}
}
\caption{Ionization fractions of \ovn, \ovin, \oviin, and \oviii
as a function of gas temperature.  The upper left panel shows results
for pure collisional ionization.  The other three panels include
photoionization by the UV and X-ray background, for gas of density
$n_H=10^{-6}\cm^{-3}$, $10^{-5}\cm^{-3}$, and $10^{-4}\cm^{-3}$,
as marked.  The corresponding overdensities are $\delta_b=6/(1+z)^3$,
$60/(1+z)^3$, and $600/(1+z)^3$ for $\Omega_b h_{70}^{2} = 0.04$.
The dot-dashed line in each panel shows $f(\ovin)+f(\oviin)+f(\oviiin)$,
which is close to unity for temperatures $3\times 10^5\K < T < 2\times 10^6\K$,
and remains high to lower temperatures when $n_H \leq 10^{-5}\cm^{-3}$.
For details of the calculations, see \cite{chen02}.
}
\end{figure}

\newpage
\begin{figure}[ht]
\centerline{
\epsfxsize=6.5truein
\epsfbox[65 450 550 720]{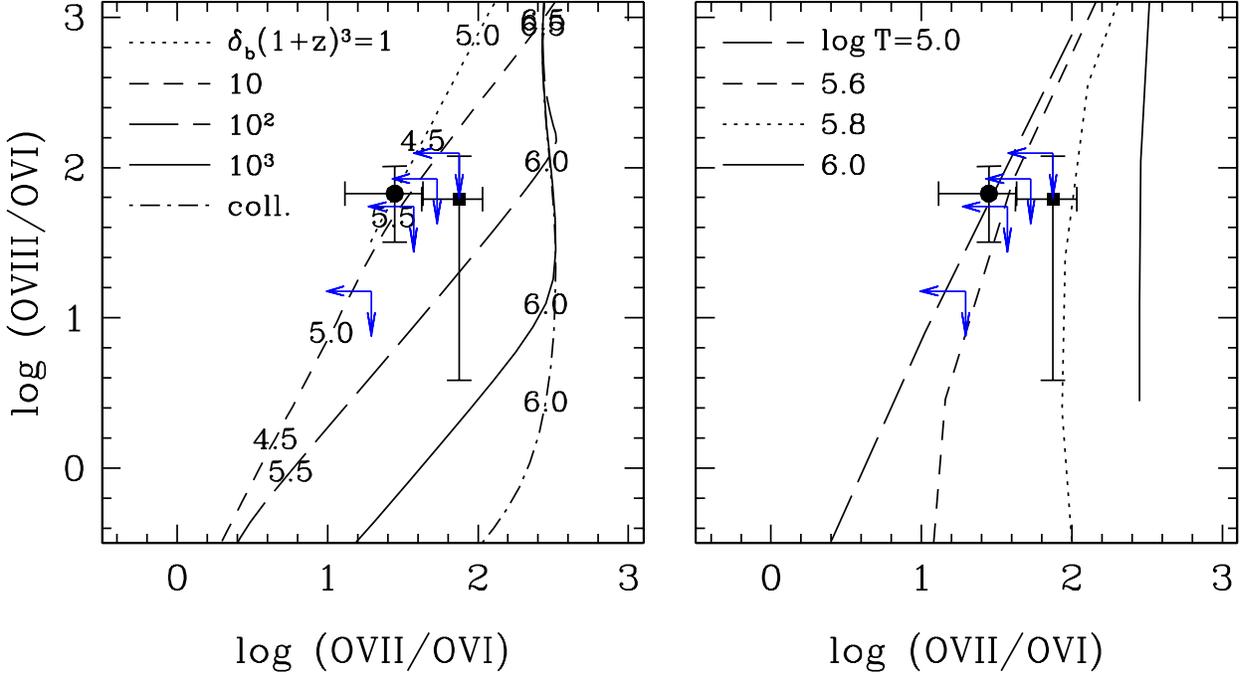}
}
\caption{Constraints on the physical state of the known \ovi
absorbers. Left: Curves show the tracks in the 
$f(\oviiin)/f(\ovin)$ vs.\ $f(\oviin)/f(\ovin)$ plane, based on
the calculations illustrated in Fig.~4.  Dotted, short-dashed, long-dashed,
and solid lines are for gas overdensities $\delta_b(1+z)^3=1$, 10, $10^2$,
and $10^3$, respectively, while the dot-dashed line represents pure
collisional ionization.  Numbers
along these curves indicate $\log T$ in degrees Kelvin. Points with
$1\sigma$ error bars show the detected systems at z2 (square) and
z6 (circle); at these redshifts $(1+z)^{-3}=0.52$ and 0.71, respectively.
Blue arrows indicate $1\sigma$ upper limits for the remaining four systems,
at z4, z1, z5, and z3 (bottom to top).  Right: Same, but with tracks
of constant temperature.
}
\end{figure}

\newpage
\begin{figure}[ht]
\psfig{figure=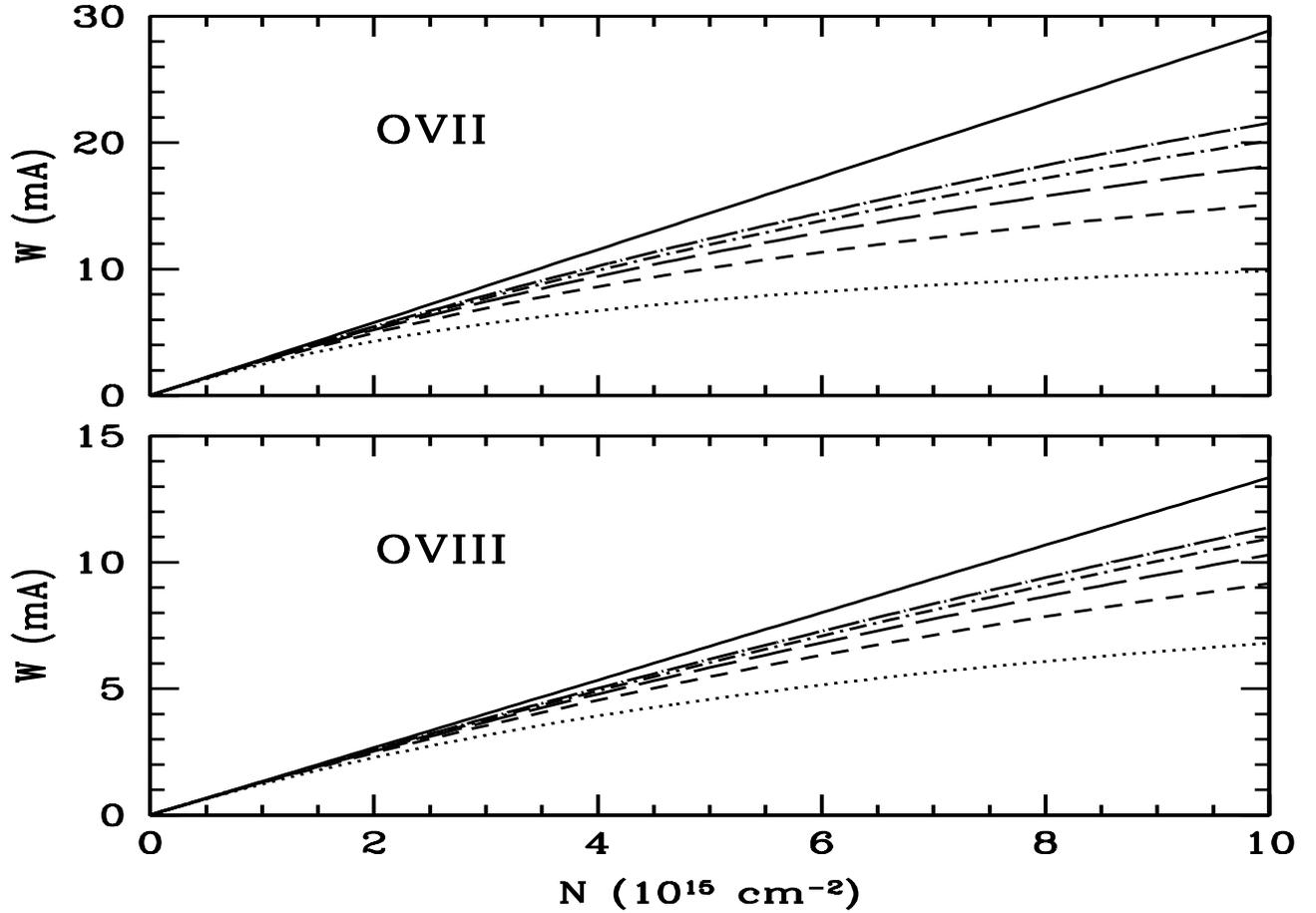,height=5in,width=7in}
\caption{Curves of growth for \ovii (top) and \oviii
(bottom). Absorption line equivalent width is plotted as a function of
column density for velocity width parameters
$b=50,$ 100, 150, 200, 250$\;\kms$, from bottom to
top. The top curve in both panels is for the optically thin limit.}
\end{figure}


\begin{thebibliography}{999}
\bibitem[Bahcall \& Peebles(1969)]{bahcall69}
Bahcall, J. N. \& Peebles, P. J. E., \apjl, 156, L7

\bibitem[Brinkman et al.(1997)]{brinkman97}
Brinkman, A.C. et al. 1997, Proc. SPIE, 3113, 181

\bibitem[Burles \& Tytler(1997)]{burles97}
Burles, S., \& Tytler, D. 1997, \aj, 114, 1330
% D/H paper with Omega_b h^2 = 0.02 at z=3.5

\bibitem[Burles \& Tytler(1998)]{burles98}
Burles, S., \& Tytler, D. 1998, \apj, 507, 732
% D/H, Q1009+2056 (z=2.5), ombh2 = 0.019 \pm 0.001

\bibitem[Cen \& Ostriker(1999a)]{cen99}
Cen, R.~\& Ostriker, J.~P.\ 1999a, \apj, 514, 1
% WHIM

\bibitem[Cen \& Ostriker(1999b)]{cen99b}
Cen, R.~\& Ostriker, J.~P.\ 1999b, \apj, 519, L109
% metallicity of the IGM

%\bibitem[Cen et al.(2001)]{cen01}
%Cen, R., Tripp, T.~M., Ostriker, J.~P., \& Jenkins, E.~B.\ 2001, \apjl, 559, L5

\bibitem[Chen et al.(2002)]{chen02}
Chen, X., Weinberg, D.~H., Katz, N., \& Dav\'{e}, R.\ 2002, \apj, submitted,
astro-ph/0203319
% X-ray forest

\bibitem[Dav\'{e} et al.(1999)]{dave99} 
Dav\'{e}, R., Hernquist, L.,
Katz, N., and Weinberg, D. H., 1999, \apj, 511, 521

\bibitem[Dav{\' e} et al.(2001)]{dave01}
Dav{\' e}, R.~et al.\ 2001, \apj, 552, 473

\bibitem[Fang \& Canizares(2000)]{fang00} 
Fang, T. and Canizares, C. R., 2000, \apj, 539, 532 

\bibitem[Fang, Bryan, \& Canizares(2002a)]{fang02a}
Fang, T., Bryan, G. L., \& Canizares, C. R. 2002a, \apj, 564, 604 

\bibitem[Fang et al.(2002b)]{fang02b}
Fang, T., Davis, D.~S.,
Lee, J.~C., Marshall, H.~L., Bryan, G.~L., \& Canizares, C.~R.\ 2002b, \apj, 
565, 86 
% Chandra observation of H1821+643

\bibitem[Ferland(1999)]{ferland99}
Ferland, G. J., 1999. {\it Hazy, a brief introduction to cloudy 94},
University of Kentucky, Physics department internal report

\bibitem[Fukugita, Hogan, \& Peebles(1998)]{fukugita98} 
Fukugita, M., Hogan, C. J., \& Peebles, P. J. E., 1998, \apj, 503, 518

\bibitem[Hellsten, Gnedin, \& Miralda-Escud\'{e}(1998)]{hellsten98}
Hellsten, U., Gnedin, N. Y., \&
Miralda-Escud\'{e}, J., 1998, \apj, 509, 56 (HGM)

\bibitem[Hernquist et al.(1996)]{hernquist96} Hernquist, L., Katz, N.,
Weinberg, D. H., \& Miralda-Escud\'{e}, J., 1996, \apj, 457, L51

\bibitem[Lockman \& Savage(1995)]{lockman95}
Lockman, F. J., \& Savage, B. D.\ 1995, \apjs, 97, 1

\bibitem[Miyaji et al.(1998)]{miyaji98} 
Miyaji, T., Ishisaki, Y.,
Ogasaka, Y., Ueda, Y., Freyberg, M. J., Hasinger, G., \& Tanaka, Y.,
1998, \aap, 334, L13. 

\bibitem[Navarro et al.(1997)]{navarro97} 
Navarro, J. F., Frenk C. S., White, S. D. M., 1997, \apj, 490, 493

\bibitem[Nicastro et al.(2002)]{nicastro02}
Nicastro, F. et al.\ 2002, ApJ, in press, astro-ph/0201058

\bibitem[Oegerle et al.(2000)]{oegerle00}
Oegerle, W.~R.~et al.\ 2000, \apjl, 538, L23

\bibitem[Perna \& Loeb(1998)]{perna98} Perna, R. \& Loeb, A., 1998,
\apj, 503, L135

\bibitem[Press et al.(1992)]{press92}
Press, W. H., Teukolsky, S. A., Vetterling, W. T., \& Flannery, B. P.\ 1992,
Numerical Recipes in Fortran, (Cambridge: Cambridge University Press)

\bibitem[Rauch \& Haehnelt(1995)]{rauch95}
Rauch, M., \& Haehnelt, M. G. 1995, \mnras, 275, L76
% pair observations -> large Omega_{IGM}

\bibitem[Rauch et al.(1997)]{rauch97}
Rauch, M., Miralda-Escud\'{e}, J., Sargent, W. L. W.,
Barlow, T., Hernquist, L., Weinberg, D. H., Katz, N., Cen, R.,
\& Ostriker, J. P., 1997, \apj, 489, 7

\bibitem[Richter et al.(2001)]{richter01} Richter, P., Savage, B. D.,
Wakker, B. P., Sembach, K. R., \& Kalberla, P. 2001, \apj, 549, 281 

\bibitem[Savage et al.(2002)]{savage02} 
Savage, B. D., Sembach, K. R., Tripp, T. M., \& Richter, P., 2002,
\apj, 564, 631
%Far Ultraviolet Spectroscopic Explorer and Space 
%Telescope Imaging Spectrograph Observations of Intervening 
%O VI Absorption Line Systems in the Spectrum of PG 0953+415

\bibitem[Sembach et al.(2001)]{sembach01} Sembach, K. R., Howk, C.,
Savage, B. D., Shull, J. M., \& Oegerle, W. R. 2001, \apj, 561, 573

\bibitem[Shull et al.(1999)]{shull99}
Shull, J.~M., Roberts, D., Giroux, M.~L., Penton, S.~V., \& Fardal, M.~A.\
1999, \aj, 118, 1450

\bibitem[Tripp \& Savage(2000)]{ts00} Tripp, T. M. and Savage, B. D.,
2000, \apj, 542, 42

\bibitem[Tripp, Savage, \& Jenkins(2000)]{tripp00} Tripp, T. M., Savage, B. D.,
\& Jenkins, E. B., 2000, \apj, 534, L1

\bibitem[Verner, Verner, \& Ferland(1996)]{verner96}
Verner, D. A., Verner,
E. M., \& Ferland, G. J., 1996, Atomic Data Nucl. Data Tables, 64, 1

\bibitem[Weinberg et al.(1997)]{weinberg97}
Weinberg, D.H., Miralda-Escud\'{e}, J., Hernquist, L., \& Katz, N., 1997,
\apj, 490, 564
% lower bound on Omega_b

\bibitem[Weymann et al.(1998)]{weymann98}
Weymann, R.~J.~et al.\ 1998, \apj, 506, 1
% key project low-z evolution paper

\end{thebibliography}
\end{document}